\documentclass[11pt]{article}
\usepackage[top=2cm,left=2cm,right=2cm]{geometry}
\usepackage{epsfig}
\usepackage{cancel}
\usepackage{caption}
\usepackage{feynmf} 
\usepackage{amssymb}
\usepackage{amsfonts}
\usepackage{epsf}
\usepackage{rotating}
\usepackage{graphicx}
\usepackage{amsmath}
\usepackage{fancyhdr}
\usepackage{subfigure}
\usepackage{graphics}
\usepackage{pstricks}
\usepackage{color}
\usepackage{frontespizio}
\usepackage{cite}

\newcommand{\cA}{\mathcal{A}}

\newcommand{\cL}{\mathcal{L}}

\newcommand{\cT}{\mathcal{T}}
\newcommand{\cS}{\mathcal{S}}

\newcommand{\sdfrac}[2]{\mbox{\small$\displaystyle\frac{#1}{#2}$}}

\IfFileExists{dsfont.sty}
{\usepackage{dsfont}
	\let\mathbb=\mathds
	\newcommand{\id}{\mathds{1}}}
{\typeout{Package dsfont.sty was not found, using alternative macros.}
	\let\mathds=\mathbb
	\newcommand{\id}{\mbox{1 \kern-.59em {\rm l}}}}

\usepackage{slashed}
\usepackage{units}
\usepackage{setspace}
\newcommand{\nn}{\nonumber}
\let\a=\alpha   \let\b=\beta      \let\d=\delta
    \let\h=\eta     
        \let\m=\mu
\let\n=\nu                 
\let\s=\sigma        
         
\let\D=\Delta
\let\d=\delta
\let\s=\sigma
\newcommand{\secref}[1]{Section~\ref{#1}}		
\newcommand{\pa}{\partial}						
\renewcommand{\a}{\alpha}
\newcommand{\bet}{\beta}
\newcommand{\del}{\delta}
\newcommand{\ka}{\kappa}
\newcommand{\lam}{\lambda}

\newcommand{\vf}{\varphi}

\newcommand{\G}{\Gamma}

\newcommand{\Th}{\Theta}

\newcommand{\La}{\Lambda}

\def\nbox#1#2{\vcenter{\hrule \hbox{\vrule height#2in
			\kern#1in \vrule} \hrule}}
\def\sq{\,\raise.5pt\hbox{$\nbox{.09}{.09}$}\,}
\def\sqb{\,\raise.5pt\hbox{$\overline{\nbox{.09}{.09}}$}\,}

\newcommand{\na}{\nabla}

\newcommand{\bea}{\begin{eqnarray}}
\newcommand{\eea}{\end{eqnarray}}
\newcommand{\be}{\begin{equation}}
\newcommand{\ee}{\end{equation}}

\newcommand{\bes}{\begin{subequations}}
	\newcommand{\ees}{\end{subequations}}
\newcommand{\lag}{\langle}
\newcommand{\rag}{\rangle}

\def\nn{\nonumber\\}

\usepackage{braket}
\numberwithin{equation}{section}

\usepackage{accents}

\allowdisplaybreaks

\begin{document}
\pagestyle{empty}
\topmargin -1.2cm
\begin{flushright} {LA-UR-17-22382}
\end{flushright}\vspace{1em}
\begin{center}
\vspace{1.5cm}
\centerline {\Large\bf
TTT in CFT: }
\vspace{2mm}
\centerline {\Large\bf Trace Identities and the Conformal Anomaly Effective Action}
\vspace{0.3cm}

\vspace{2 cm}

{\bf  $^{(1)}$Claudio Corian\`o, $^{(1)}$Matteo Maria Maglio and  $^{(2)}$Emil Mottola\\}

\vspace{0.5cm}

{$^{(1)}$Universit\`a del Salento and INFN Lecce, \\
Via Arnesano, 73100 Lecce, Italy}

\vspace{0.5cm}

{$^{(2)}$Theoretical Division, T-2\\
MS B285\\
Los Alamos National Laboratory\\ 
Los Alamos, NM 87545 USA\vspace{5mm}}
\end{center}
\date{\version}
\begin{abstract}

Stress-energy correlation functions in a general Conformal Field Theory (CFT) in four dimensions are described in a fully covariant approach, as metric variations of the quantum effective action in an arbitrary curved space background field. All Conservation, Trace and Conformal Ward Identities (CWIs), including contact terms, are completely fixed in this covariant approach. The Trace and CWIs are anomalous. Their anomalous contributions may be computed unambiguously by metric variation of the exact 1PI quantum effective action determined by the conformal anomaly of $\braket{T^{\mu\nu}}$ in $d = 4$ curved space. This action implies the existence of massless propagator poles in three and higher point correlators of $T^{\mu\nu}$ . The metric vari- ations of the anomaly effective action in its local form in terms of a scalar conformalon field are carried out explicitly for the case of the correlator of three CFT stress-energy tensors, and the result is shown to coincide with the algebraic reconstruction of $\braket{TTT}$ from its transverse, tracefree parts, determined inde- pendently by the solution of the CWIs in d dimensional flat space in the momentum representation. This demonstrates that the specific analytic structure and massless poles predicted by the general curved space anomaly effective action are in fact a necessary feature of the exact solution of the anomalous CWIs in any $d = 4$ CFT.
\end{abstract}

\newpage
\pagestyle{plain}
\setcounter{page}{1}

\section{Introduction} 

In the absence of a complete description of gravitation in four dimensions consistent with quantum theory, quantum effects in gravitational fields may be studied by means of the correlation functions of the energy-momentum-stress tensor $T^{\m\n}$, which couples to the curved spacetime metric $g_{\m\n}$. The renormalized expectation value $\lag T^{\m\n}(x)\rag_g$ in metric backgrounds contains interesting physical information about quantum effects 
in strong gravitational fields, such as those of black holes and cosmological spacetimes. If used as a source for Einstein's equations, the one-point function 
$\lag T^{\m\n}(x)\rag_g$ provides the basis for a mean field approach to semi-classical gravity, upon which most present knowledge of quantum effects and 
interactions in gravity, sparse as they are, are based \cite{Birrell:1982ix}. The two-point function $\lag T^{\m_1\n_1}(x_1) T^{\m_2\n_2}(x_2)\rag_g$ contains 
additional information about vacuum polarization effects in gravitational fields, and enters directly into the linear response and stability of solutions 
of the semi-classical Einstein eqs. to small metric disturbances \cite{AndMolMott:2003}. 

Higher point correlation functions of multiple stress tensors $\lag T^{\m_1\n_1}(x_1) T^{\m_2\n_2}(x_2)T^{\m_3\n_3}(x_3)...\rag_g$, containing 
additional non-trivial information about quantum correlations in gravitational fields are less well studied. The three-point functions of $T^{\m\n}$ with two vector gauge currents $\lag TJJ\rag$, very much analogous to the familiar chiral anomaly amplitude $\lag AJJ\rag$ of one axial and two vector 
currents, have been investigated in their exact \cite{Giannotti:2008cv,Armillis:2009pq} and broken phases \cite{Coriano:2011ti} only relatively recently. 
Such correlators may be relevant in classically conformal invariant field theories (CFTs), and in phenomenological scenarios envisioning possible 
conformal extensions of the Standard Model with a composite Higgs/dilaton scalar \cite{Coriano:2011ti,Bandyopadhyay:2016fad}. CFTs are 
of interest in their own right, both for the important role they play at critical points in renormalization group flows and possible application 
to a wide variety of many-body quantum systems, including the conformal bootstrap program, which has seen a recent resurgence of
interest \cite{PolSim}.

Even massless or classically conformal invariant theories must break both scale and conformal invariance at the quantum level. This breaking 
of conformal invariance is manifest by the non--zero trace $\lag T^{\m}_{\ \m}(x)\rag_g$ of the energy-momentum tensor in a background gauge or gravitational 
field. The general form of the trace anomaly of the one-point function has been known for some time \cite{Capper:1974ic}, \textit{cf.} Eq. \eqref{trace}. This conformal anomaly has been shown to lead to a full three-point correlator $\braket{T^{\m\n}J^\a J^\b}$ in massless QED that possesses a massless scalar pole in momentum space \cite{Giannotti:2008cv}. 
The three-point correlation function of stress tensors in QCD exhibits such a massless pole as well \cite{Armillis:2009pq}. In the Standard Model the same 
massless scalar pole is inherited by the dilation current correlator \cite{Coriano:2011ti,Coriano:2011zk}, which appears also in supersymmetric theories
\cite{Coriano:2014gja}. These results naturally raise the question if the presence of a massless scalar pole in the three-point $\braket{TTT}$ correlator are a general necessary and inevitable consequence of the anomalous conformal Ward Identities of any CFT in $d = 4$ dimensions. This is the main question we address and answer in the affirmative in this paper.

Three-point $T^{\mu\nu}$ CFT correlators were analyzed in position space with Euclidean flat metric in \cite{Osborn:1993cr}.
However the presence or absence of massless poles is associated with behavior on the light cone, which is most clearly evident in a momentum space representation with a Lorentzian metric. The technical tools for solving Conformal Ward Identities (CWIs) 
in the momentum representation have been developed only in the past few years \cite{Coriano:2012wp}. The three-point function 
$\lag T^{\m_1\n_1}T^{\m_2\n_2}T^{\m_3\n_3}\rag$ has been found recently by solving the CWIs in $d$-dimensional flat Minkowski space
in the momentum representation \cite{Bzowski:2013sza}. In this approach the CWIs are treated as non-anomalous in general
$d$-dimensions, but when solved for general $d$, exhibit $1/(d-4)$ divergences in the limit $d\rightarrow 4$. These divergences
require counterterm subtractions and a somewhat involved reconstruction of the trace parts from the transverse, tracefree parts,
which amounts to reconstructing the anomaly and anomalous contributions to $\lag TTT\rag$ {\it a posteriori}. The momentum space massless 
scalar $1/k^2$ pole in the final renormalized correlator is somewhat concealed by the technicalities of this method.

In this paper we take a quite different approach, focusing specifically on the contributions to the CWIs directly required by the conformal anomaly, and directly in four dimensions, by making use of the known form of the conformal anomaly and its one-particle irreducible (1PI) effective action in $d = 4$ curved spacetime. This starting point already incorporates the necessary regularization and renormalization of $\braket{T^{\mu\nu}}_g$ consistent with general covariance. Hence no further regularization, dimensional or otherwise is required, and the momentum space pole structure associated with the anomaly may be examined directly.

The form of the conformal anomaly determines the 1PI quantum effective action for a CFT in a general curved space background \cite{Riegert:1987kt,Antoniadis:1992xu}, up to conformally invariant and purely local terms. Any effects of the anomaly are determined by this anomalous effective action by successive variation with respect to the arbitrary background metric $g_{\mu\nu}$ , after which the metric may be set to the flat Minkowski metric $\eta_{\mu\nu}$. In particular, the presence of the massless scalar pole in the third metric variation of this effective action shows conclusively that the massless $1/k^2$ pole is necessarily associated with the conformal anomaly, directly in $d = 4$.

We compare this direct calculation from the anomaly effective action with the method developed in Refs. \cite{Bzowski:2013sza} 
for reconstructing the trace contributions from the transverse, tracefree parts of $\lag T^{\m_1\n_1}T^{\m_2\n_2}T^{\m_3\n_3}\rag$ by
dimensional regularization of $d$-dimensional flat space amplitudes, and show that these two quite different methods yield precisely the same result 
for the anomalous trace contributions in the momentum representation. Thus, the massless scalar $1/k^2$ pole is necessarily present in general CFT three-point CFT correlator, exactly as predicted by the anomaly effective action, and as a direct consequence of the anomalous CWIs it obeys. 

The outline of the papers is as follows. The covariant quantum action in curved space is defined in \secref{Sec:QAction}, and the method of derivation from it of all the relevant Ward Identities for the $n$-point correlators $\lag T^{\m_1\n_1}(x_1) T^{\m_2\n_2}(x_2)\dots T^{\m_n\n_n}(x_n)\rag_g$ given in Secs.\,ref{Sec:ConsWI}-\ref{Sec:CWI}, with explicit results for the CWIs $n=2,3$. In \secref{Sec:Reconst} the approach of \cite{Bzowski:2013sza} for reconstructing the trace parts of $\lag TTT\rag$ from its transverse, traceless parts is reviewed. The effective action for a general CFT decomposed into its Weyl invariant, anomalous and local parts, and the anomaly effective action in $d=4$, in both its non-local and local form, as well as the Wess-Zumino 
consistency condition it satisfies is summarized in \secref{Sec:AnomAct}. In \secref{Sec:VarAct} this curved space anomaly 1PI effective action is developed in a Taylor series to third order around flat space in \secref{Sec:AnomTTT}, and its contribution to $\lag TTT\rag$ is explicitly computed in momentum space and shown to be identical to the reconstruction method of \cite{Bzowski:2013sza}. Both methods are shown to yield precisely the same $1/k^2$ massless scalar pole and tensor structure required by the conformal anomaly. We conclude in \secref{Sec:Discuss} with a Summary and Discussion of the implications of the existence of a massless scalar pole as indicative of an effective scalar degree of freedom in low energy macroscopic gravity not contained in classical General Relativity, and sketch some directions for future work.

\section{\hspace{-3mm}1PI action in curved space and $\lag TTT\!\dots\rag$ correlators}
\label{Sec:QAction}

The symmetric energy-momentum-stress tensor $T^{\m\n}(x)$ coupling to the gravitational metric field 
$g_{\m\n}(x)$ is defined by the metric variation
\be
T^{\m\n}(x)\equiv \frac{2}{\sqrt{-g(x)}}\frac{\ \del{\rm S_{cl}}[\Phi, g] }{\del g_{\m\n}(x)}
\ee
where ${\rm S_{cl}} [\Phi, g]$ is the classical action of the fields, denoted generically by $\Phi = \{\Phi_i\}$. Here ${\rm S_{cl}}[\Phi, g]$ is viewed 
as a functional also of the general curved spacetime background metric denoted symbolically by $g$, and $\sqrt{-g}(x)\equiv\sqrt{-\det g_{\mu\nu}(x)}$.

When the matter/radiation fields $\Phi$ in $\rm S_{cl}$ are quantized, $T^{\m\n}$ becomes an operator and the fundamental quantities 
of interest are the $n$-point quantum correlation or vertex functions
\be
\cS_n^{\m_1\n_1\dots\m_n\n_n}(x_1,\ldots,x_n;g) = i^{n-1}\big\lag \cT^* \big\{T^{\m_1\n_1}(x_1)\dots T^{\m_n\n_n}(x_n)\big\}\big\rag_g\,,\qquad
x_i \neq x_j \quad \forall \quad i \neq j
\label{correfn}
\ee
of covariant $\cT^*\{\dots\}$ time-ordered product of operators $T^{\m_i\n_i}(x_i)$ at distinct spacetime points $x_i$ in the general 
background metric $g_{\m\n}(x)$. These correlation functions (\ref{correfn}) are well-defined when no two of the spacetime points coincide. When any of the spacetime points in (\ref{correfn}) do coincide, the correlation functions acquire contact terms which must be treated with some care. 

The contact terms are most conveniently handled by defining the correlators in (\ref{correfn}) in a generally covariant way as variations of the one-particle irreducible (1PI) exact quantum effective action $\cS[g]$ defined formally by the functional integral
\be
\exp \big\{i\, \cS[g]\big\} \equiv \int [d \Phi] \exp\big\{i\, {\rm S_{cl}}[\Phi, g]\big\}
\label{Sexact}
\ee
over all the matter/radiation fields (at all scales) in the fixed but arbitrary background metric. By expanding this 1PI quantum action functional in a Taylor series
\bea
&&\cS[g+h] = \cS[g] + \\
&&\hspace{-5mm}\sum_{n=1}^{\infty} \frac{1}{2^n\,n!} \int d^4x_1\dots d^4x_n \, \sqrt{-g(x_1)} \dots \sqrt{-g(x_n)}\, \cS^{\m_1\n_1\dots \m_n\n_n}_n(x_1, \dots,x_n;g)\,
h_{\m_1\n_1}(x_1)\dots h_{\m_n\n_n}(x_n)   \nonumber
\label{Tayexp}
\eea
around a given background metric, it becomes clear that the coefficients of this Taylor series $\cS_n$, defined by the multiple functional variations
\be
\cS_n^{\m_1\n_1\dots\m_n\n_n}(x_1,\ldots,x_n;g) \equiv \frac{2^n}{\sqrt{-g(x_1)}\dots\sqrt{-g(x_n)}}\ 
\frac{\del^{n}\cS[g]}{\del g_{\m_1\n_1}(x_1)\dots \del g_{\m_n\n_n}(x_n)}
\label{Tn}
\ee
generate precisely the correlation functions (\ref{correfn}) in the case that none of the spacetime points coincide. Moreover this fully covariant definition of the $T^{\m\n}$ correlation functions also determines all the contact terms through the Ward Identities they satisfy, as we shall show presently. If Feynman boundary conditions are specified on the functional integral (\ref{Sexact}), then $\cS[g]$ is the {\it in-out} effective action, 
which is the generating functional of the covariant $\cT^*$ time ordered {\it in-out} connected correlation functions (\ref{correfn}). Other boundary conditions (such as {\it in-in}) may also be imposed on (\ref{Sexact}) to obtain the corresponding correlation functions.

Let us remark that it is important to distinguish the 1PI full quantum effective action $\cS[g]$ in (\ref{Sexact}), defined in the gravitational 
sector by integrating out {\it all} non-gravitational matter/radiation fields at {\it all} scales, from the Wilsonian effective action in common use in high energy, 
nuclear and condensed matter physics applications. The Wilson effective action is defined by integrating out only heavy degrees of freedom above 
some mass scale $M$, and expressing the result in a {\it local} derivative expansion up to some finite order in $1/M$. Unlike this approximate Wilson 
effective action, the {\it exact} 1PI action $\cS[g]$, defined formally by (\ref{Sexact}), is generally quite non-local, as it contains the full effects of the 
quantum fluctuations of light or massless fields, and is thus applicable to Conformal Field Theories that are the subject of this paper. 

\section{Conservation ward identities}
\label{Sec:ConsWI}
\subsection{Position space}
For $n=1$ the covariant definition of the one-point function
\be
\cS^{\m\n}_1(x;g) \equiv \big\lag T^{\m\n}(x)\big\rag_g=\frac{2}{\sqrt{-g(x)}}\ \frac{\del \cS [g]}{\del g_{\m\n}(x)} 
\label{T1def}
\ee
is the renormalized expectation value of $T^{\m\n}(x)$, where we assume that renormalization counterterms have already been taken account of,
and $\cS [g]$ is the finite renormalized 1PI effective action. If the renormalization procedure respects general coordinate invariance, then
the expectation value (\ref{T1def}) will satisfy the fundamental Ward Identity of covariant conservation 
\be
^{(g)}\na_\m \big\lag T^{\m\n}(x)\big\rag_g =0
\label{EMSTcons}
\ee
with $^{(g)}\na_\m$ the covariant derivative in the general background metric $g_{\m\n}(x)$. 

The conservation Ward Identity (\ref{EMSTcons}) may be rewritten in the form
\bea
&&\sqrt{-g}\  ^{(g)}\na_\m \big\lag T^{\m\n}(x)\big\rag_g = 
\pa_{\n} \left(\sqrt{-g}\, \cS_1^{\m\n}\right)  + \G^\m_{\ \lam\n} \left(\sqrt{-g}\, \cS_1^{\lam\n}\right) \nn
&&\hspace{1.5cm}=2\,\frac{\pa}{\pa x^\n}\left(\frac{\del \cS [g]}{\del g_{\m\n}(x)}  \right)  
+ 2\,\G^\m_{\ \,\lam\n}(x) \left(\frac{\del \cS [g]}{\del g_{\lam\n}(x)} \right)  = 0
\label{onept}\eea
by making use of the Christoffel connection 
\be
\G^\m_{\ \,\lam\n}(x) = \frac{1}{2}\Big[\,g^{\m\a} \,\Big(\! - \pa_\a g_{\lam\n} +  \pa_\lam g_{\n\a} +  \pa_\n g_{\a\lam} \Big)\Big]_x
\label{Chris}
\ee
for the general background metric $g_{\m\n}(x)$, with the subscript x indicating the spacetime point at which the expression within brackets is to be evaluated. The last form of (\ref{onept}) is the most convenient one for additional
successive variations with respect to $g_{\m\n}(x)$, by which conservation Ward Identities for the higher point correlation functions may be derived. 

For the two-point function, we take one additional variation of the fundamental relation (\ref{onept}), and note that in flat space in the absence of boundaries (and in Cartesian coordinates), we may set both the expectation value $\big\lag T^{\m\n}(x)\big\rag_\eta$ and $\G^\m_{\ \,\lam\n}(x)=0$, thus obtaining
\be
\pa_{\n_1} \, \cS_2^{\m_1\n_1\m_2\n_2}(x_1, x_2;\eta) = 0
\label{twoptcons}
\ee
for $n=2$. Likewise for the three-point function we take two metric variations of (\ref{onept}), and make use of the first variation of the Christoffel symbol
\be
2\ \frac{\del \G^{\m_1}_{\ \ \lam_1\n_1}(x_1)}{\del g_{\m_2\n_2}(x_2)}\bigg\vert_{g=\eta} = \eta^{\m_1\a_1}
\Big\{ - \d_{\lam_1}\, ^{\!\!\!(\m_2} \d_{\ \ \n_1}^{\n_2)}\pa_{\a_1} + \del_{\n_1}^{\ (\m_2}\d^{\n_2)}_{\ \ \a_1} \pa_{\lam_1}
+ \d_{\a_1}^{\ (\m_2}\d^{\n_2)\!\!\!}\,_{\lam_1} \pa_{\n_1}\Big\}\,\d^4(x_1 - x_2)
 \label{varChris}
\ee
where the derivatives of the $\d$-functions in (\ref{twoptcons}), (\ref{varChris}) and (\ref{threeptcons}) below are with respect to $x_1$, and parentheses around indices denotes symmetrization with respect to those indices. Then taking account of the normalization in (\ref{Tn}) we obtain
\bea
&&\hspace{-1cm}\pa_{\n_1} \,\cS_3^{\m_1\n_1\m_2\n_2\m_3\n_3}(x_1, x_2, x_3;\eta) = \nn
&&\left\{ \Big(\d_{\lam_1}\, ^{\!\!\!(\m_2} \d_{\ \ \n_1}^{\n_2)} \eta^{\m_1\a_1}\pa_{\a_1} 
- 2\, \eta^{\m_1(\m_2}\del^{\n_2)\!\!\!}\,_{(\lam_1} \pa_{\n_1)}\Big)\d^4(x_1 - x_2) \right\}\cS_2^{\lam_1\n_1\m_3\n_3}(x_1,x_3;\eta)\nn
&&\hspace{-5mm} +  \left\{\Big( \d_{\lam_1}\, ^{\!\!\!(\m_3} \d_{\ \ \n_1}^{\n_3)} \eta^{\m_1\a_1}\pa_{\a_1} 
- 2\, \eta^{\m_1(\m_3}\del^{\n_3)\!\!\!}\,_{(\lam_1} \pa_{\n_1)}\Big)\,\d^4(x_1 - x_3) \right\}\cS_2^{\lam_1\n_1\m_2\n_2}(x_1,x_2;\eta) 
\label{threeptcons}
\eea
for the conservation Ward Identity of the three-point function of stress tensors in position space, and in the flat space limit, setting $g_{\m\n}(x) = \eta_{\m\n}$ after the variations.

This variational procedure could be continued indefinitely to obtain the conservation Ward Identities of any $n$-point function of $T^{\mu\nu}(x)$, with the conservation identities of the $n$-point functions of $T^{\mu\nu}(x)$ involving $\delta$-function contact terms multiplying the $n-1$ and lower point functions. This procedure shows that the symmetric and covariant definition (\ref{Tn}) of the $n$-point functions, where all the $\sqrt{-g}$ factors 
are placed to the left of the metric variation of the effective action, yields conservation Ward Identities where all the contact terms involving derivatives 
of $\d$-functions at coincident points are determined in a completely general, geometric and theory-independent way from the variations of the $\G^\m_{\ \,\lam\n}(x)$ symbol in the basic covariant conservation identity for the one-point function (\ref{onept}). Thus with the definition (\ref{Tn}) one never encounters explicit variations of $T^{\m\n}(x)$  itself at coincident spacetime points (`seagulls'), whose form may depend upon the specific underlying field theory being considered, and which may introduce non-covariant contact terms that would complicate the analysis considerably. This is a significant technical and conceptual advantage of a fully covariant approach, in a curved space background.

\subsection{Momentum space}

The $n$-point functions of $T^{\m_i\n_i}(x_i)$ in flat spacetime may be Fourier transformed by 
\bea
&&\hspace{-3mm}\int\! d^4x_1\dots d^4x_n\, e^{i p_1\cdot x_1 + \dots + i p_n\cdot x_n} \cS_n^{\m_1\n_1\dots\m_n\n_n}(x_1,\ldots,x_n;\eta)\nn
&& \equiv
(2\pi)^4 \d^4(p_1+ \dots + p_n)\,\tilde \cS_n^{\m_1\n_1\dots\m_n\n_n} (p_1, \dots, p_n)
\label{Fourtrans}
\eea
where the reduced correlation functions $\tilde \cS_n$ are defined only on the subspace of $(p_1 + \dots + p_n)^{\m} = 0$ 
determined by the $\del$-function of four-momentum conservation. Then the conservation Ward Identities (\ref{twoptcons}) and
(\ref{threeptcons}) for the reduced correlation functions become in momentum space 
\be
(p_1)_{\n_1}\, \tilde \cS_2^{\m_1\n_1\m_2\n_2}(p_1, \bar p_2)=0
\label{twoptmom}
\ee
evaluated at $\bar p_2^\m = - p_1^\m$, and
\bea
&&\hspace{-1cm}(p_1)_{\n_1} \tilde \cS_3^{\m_1\n_1\m_2\n_2\m_3\n_3} (p_1, p_2, \bar p_3) =
-(p_2)^{\m_1}\tilde \cS_2^{\m_2\n_2\m_3\n_3} (p_1+p_2, \bar p_3)  -(\bar p_3)^{\m_1}\tilde \cS_2^{\m_2\n_2\m_3\n_3} (p_2, p_1 + \bar p_3) \nn
&&+\, 2\, (p_2)_{\a}\,\eta^{\m_1(\n_2}\tilde \cS_2^{\m_2)\a\m_3\n_3} (p_1+p_2, \bar p_3)
+\, 2\, (\bar p_3)_{\a}\, \tilde \cS_2^{\m_2\n_2\a (\n_3} (p_2, p_1 + \bar p_3)\,\eta^{\m_3)\m_1}
\label{threeptmom}
\eea
evaluated at $\bar p_3^{\m} = -(p_1 + p_2)^{\m}$, for the $n=2$ and $n=3$ point functions respectively.

The conservation Ward Identities for the four-point and higher $n$-point correlation
functions may be derived in a similar way by repeated variations of the fundamental covariant conservation identity (\ref{onept}) with respect
to the metric, setting the metric $g_{\m\n}(x)$ to the flat one $\eta_{\m\n}$, and then Fourier transforming the result, evaluating on the
condition of momentum conservation $(p_1 + \dots + p_n)^{\m} =0$. Remembering this condition, particularly in differentiating later with 
respect  to the $p_i$, so that there are in fact only $n-1$ independent momenta in $\tilde \cS_n$, we shall drop the overbar notation henceforth.
Eq.~(\ref{twoptmom}) shows that the two-point function is purely transverse, whereas the three-point and higher $n$-point functions
contain both transverse and longitudinal terms, the latter non-vanishing upon contraction with one of the $p_j$, as in (\ref{threeptmom}).

\section{Trace ward identities for CFTs}
\label{Sec:TraceWI}
\subsection{Position space}
As already remarked, in order to be completely well-defined, the functional integral in (\ref{Sexact}) must be regularized and renormalized, 
and local covariant counterterms up to dimension four (in $d=4$ spacetime dimensions) must be added to the quantum effective action 
in order to remove its ultraviolet divergences. If general covariance is respected by this procedure the conservation Ward Identities of the 
previous section remain valid for the renormalized effective action and its variations. However, as is well-known, any renormalization 
procedure respecting coordinate invariance, such as the Schwinger-deWitt heat kernel method or dimensional regularization,
necessarily introduces a scale, leading to anomalous contributions to the trace identities of the correlation functions (\ref{correfn}).

These scale violations are determined by the general form of the conformal trace anomaly in curved space.
The fundamental identity is again that of the one-point function, which takes the general form
\be
\big\lag T^{\m}_{\ \m}(x)\big\rag_g \equiv g_{\m\n} \big\lag T^{\m\n}(x)\big\rag_g = T_{cl} +  b\, C^2 +
 b' \big(E - \tfrac{2}{3}\sq R\big) + b'' \sq R + \sum_i \beta_i \,\cL_i
\label{trace}
\ee
in a general curved background. Here $T_{cl}$ is the trace expected from the non-invariance of the classical action itself
\be
T = \frac{2\,g_{\m\n}(x)}{\sqrt{-g(x)}}\left\lag\frac{\ \del{\rm S_{cl}}[\Phi, g] }{\del g_{\m\n}(x)}\right\rag_{\!g}
\label{cltrace}
\ee
absent any anomalies, and the additional finite terms in (\ref{trace}) are the quantum anomalous contributions. 
These are given in terms of the dimension-four curvature invariants 
\bes
\bea
&&E \equiv\, ^*\hskip-.5mm R_{\m\a\n\b}\,^*\hskip-.5mm R^{\m\a\n\b} =
R_{\m\a\n\b}R^{\m\a\n\b}-4R_{\m\n}R^{\m\n} + R^2 \\
&&C^2\equiv C_{\m\a\n\b}C^{\m\a\n\b} = R_{\m\a\n\b}R^{\m\a\n\b} -2 R_{\m\n}R^{\m\n}  + \tfrac{1}{3}R^2
\eea\label{EFdef}\ees
which are the Euler-Gauss-Bonnet invariant and square of the Weyl conformal tensor respectively (with $R_{\m\a\n\b}$ the Riemann curvature tensor,
$^*\hskip-.5mm R_{\m\a\n\b}$ its dual, $R_{\m\n} = g^{\a\b} R_{\m\a\n\b}$ and $R = g^{\m\n}R_{\m\n}$ the Ricci tensor and Ricci scalar
respectively).

Additional dimension-four local invariants denoted by $\cL_i$ in (\ref{trace}) may also appear in the general form of the trace anomaly, if there are couplings 
to additional background fields. For example in the case of massless fermions coupled to a gauge field, there are contributions from the scalar invariants 
$\cL_F= F_{\m\n}F^{\m\n}$ of electromagnetism or from the strong or electroweak non-abelian gauge fields $\cL_G = {\rm Tr} (G_{\m\n}G^{\m\n})$, 
with coefficients $\b_{_F}$ or $\b_{_G}$ determined by the $\b$-function of the corresponding gauge coupling. The anomalous contributions 
must take the form of this sum of local dimension-four invariants in (\ref{trace}) for any local and covariant quantum field theory in $d=4$ spacetime 
dimensions, the only dependence upon the particular theory residing in the values of $b, b', b'', \b_i$,  which are dimensionless coefficients in units of 
$\hbar$. 

For a general discussion of CFTs in the flat spacetime limit, with no background gauge fields, we require only the minimal coupling to the spacetime 
metric, necessary to obtain the covariant definition of the three-point correlator $\cS_3^{\m_1\n_1\m_2\n_2\m_3\n_3}(x_1, x_2, x_3;\eta)$ 
by metric variations, as in (\ref{Tn}). Hence for this application to CFT we set the explicit trace term of (\ref{trace}) and (\ref{cltrace}) $T_{cl}=0$, and also 
drop the possible $\cL_i$ anomalous terms in (\ref{trace}) depending upon background fields other than the metric.

We note also that the $\sq R$ term in (\ref{trace}) corresponds to a local action, since
\be
2\, g_{\m\n}(x)\,\frac{\del }{\del g_{\m\n}(x)}\, \int dx\, \sqrt{-g} \,R^2 = -12\, \sq R
\label{varRsq}
\ee
and hence may be removed by such a local counterterm in the action with finite $b''$ coefficient. Thus it is not part of the true anomaly in (\ref{trace})
which cannot be so expressed as the variation of a local covariant action functional of the metric alone, and we may take $b''$ to be 
zero, restoring an arbitrary local $\int R^2$ term to the action at the end of the analysis. This leaves only the Weyl squared $b$ and Euler-Gauss-Bonnet
$b'$ terms in (\ref{trace}), the coefficients of which are also sometimes relabelled as $c$ and $a$ respectively in the recent literature.

As in the case of the conservation Ward Identities, the trace identities for the $n$-point functions may be derived by successive
variation of the fundamental trace identity of the one-point function. Retaining then only the $b$ and $b'$ terms, and rewriting (\ref{trace}) 
in the form
\be
2\, g_{\m\n}(x)\,\frac{\del \cS[g]}{\del g_{\m\n}(x)} =\cA \equiv \sqrt{-g}\, \Big\{ b\, C^2 + b' \big(E - \tfrac{2}{3}\sq R\big) \Big\}\,,
\label{Adef}
\ee
we vary once again with respect to the metric to obtain
\be
\eta_{\m_1\n_1}\cS_2^{\m_1\n_1\m_2\n_2} (x_1, x_2) = 2\, \frac{\del \cA(x_1)}{\del g_{\m_2\n_2}(x_2)}\bigg\vert_{flat}
\label{twoptrace}
\ee
for the two-point function, and 
\bea
&&\eta_{\m_1\n_1}\cS_3^{\m_1\n_1\m_2\n_2\m_3\n_3}(x_1, x_2 ,x_3) =
- 2\Big\{\d^4 (x_1-x_2) + \d^4(x_1-x_3)\Big\} \cS_2^{\m_2\n_2\m_3\n_3}(x_2,x_3)\nn
&& \qquad \qquad + \, 4 \, \frac{\del^2\cA(x_1)}{\del g_{\m_2\n_2}(x_2)\del g_{\m_3\n_3}(x_3)}\bigg\vert_{flat}
\label{threeptrace}
\eea
for the three-point function, evaluating both at the flat space Minkowski metric $\eta_{\m\n}$.

\subsection{Momentum space}

Fourier transforming these results above gives the trace identity for the two-point function 
\be
\eta_{\a_1\b_1} \, \tilde \cS_2^{\a_1\b_1\m_2\n_2}(p_1, p_2) =  2\, \tilde\cA_1^{\m_2\n_2} (p_2)
\label{twoptr}
\ee
in the momentum representation (again evaluated at $p_2^\m = - p_1^{\m}$), and for the trace identity of the three-point function 
\bea
&&\hspace{-2cm}\eta_{\a_1\b_1}\tilde \cS_3^{\a_1\b_1\m_2\n_2\m_3\n_3}(p_1, p_2 ,p_3) \nn
&&= - 2\,\tilde\cS_2^{\m_2\n_2\m_3\n_3}(p_1 +p_2,p_3) -2\, \tilde\cS_2^{\m_2\n_2\m_3\n_3}(p_2,p_1 + p_3) 
 + \, 4 \, \tilde \cA_2^{\m_2\n_2\m_3\n_3} (p_2,p_3)
\label{threeptr}
\eea
where 
\bea
&&\hspace{-1.5cm} (2\pi)^4 \,\d^4 (p_1+ \dots + p_{n+1})\, \tilde \cA_{n}^{\m_2\n_2\dots\m_{n+1}\n_{n+1}}(p_2,\dots ,p_{n+1}) \nn
&& \quad \equiv \int d^4x_1\dots d^4x_{n+1} \ e^{i p_1\cdot x_1 + \dots + i p_{n+1}\cdot x_{n+1}} \, 
\frac{\del^{n}\cA(x_1)}{\del g_{\m_2\n_2}(x_2)\dots \del g_{\m_{n+1}\n_{n+1}}(x_{n+1})}\Big\vert_{flat}
\label{Avardef}
\eea
is the Fourier transform of the $n$th variation of the anomaly in the flat space limit. The locality of $\cA(x_1)$ implies 
that $\tilde \cA_{n}^{\m_2\n_2\dots\m_{n+1}\n_{n+1}}$ is a polynomial with only positive powers of the $p_j$, containing no 
$1/p_j^2$ pole terms or logarithms. We note also that if $b'' \neq 0$, the $b'' \sq R$ term may easily be included in $\cA$, 
giving an additional local contribution to $\tilde \cA_{n}$.

The trace identity (\ref{threeptr}) for the three-point function contains two terms involving $\tilde \cS_2$ which would
usually be considered `non-anomalous,' despite the fact that $\tilde \cS_2$ itself implicitly depends upon the first variation of
$\tilde \cA_1$ through (\ref{twoptr}). In addition (\ref{threeptr}) contains the explicitly anomalous last term involving the second 
variation $\tilde \cA_2$. Clearly one may take additional variations of the fundamental trace identity (\ref{Adef}) with respect to 
the metric in order to obtain trace identities for higher $(n+1)$-point functions, and this pattern will continue with a hierarchy of trace identities, 
each implicitly dependent upon the $(n-1)$th and lower variations of the anomaly through the $n$-point and lower point functions 
in the apparently `non-anomalous' part of its trace Ward Identities, while at each order a new explicitly anomalous term involving the
$n$th variation of the trace anomaly enters.

\section{Conformal Ward Identities}
\label{Sec:CWI}

\subsection{Position space}
When the underlying QFT is classically conformally invariant, the $n$-point functions of the stress tensor satisfy additional Ward Identities,
namely Conformal Ward Identities (CWIs). In a covariant geometric approach these are generated by the (partial) conservation of the conformal currents
\be
J^{\m}_{(K)}(x) \equiv K_\n(x)\, T^{\m\n}(x)
\label{Jxi}
\ee
where $K_{\m}(x)$ are Conformal Killing Vectors (CKVs), defined by solutions of the Conformal Killing equation
\be
\pa_{(\m}K_{\n)} \equiv \sdfrac{1}{2} \,\big(\pa_{\m}K_{\n} + \pa_{\n} K_{\m}\big) = \sdfrac{1}{d} \, \eta_{\m\n} \,\left(\pa\cdot K\right)
\label{CKeq}
\ee
for $d$ dimensional Minkowski space. The conservation of $J^{\m}_{(K)}$ then follows from the conservation of $T^{\m\n}$, provided that the trace $\eta_{\m\n}T^{\m\n} = 0$. This traceless condition is satisfied for a CFT, at the {\it classical} level, absent the anomaly.
Otherwise there are violations of the conservation of $J^{\m}_{(K)}$, resulting from the conformal anomaly, which one can
calculate by inserting the current (\ref{Jxi}) in the $n$-point stress tensor correlator. Thus we are led to consider total divergences of the form
\bea
&&\hspace{-1.2cm}\frac{\pa}{\pa x^\n} \Big\{ K_{\m}(x)\, \cS_{n+1}^{\m\n\m_1\n_1\dots\m_n\n_n}(x, x_1,\dots, x_n) \Big\}\nn
&&\hspace{-1cm} =K_{\m}(x)\,\pa_{\n}  \cS_{n+1}^{\m\n\m_1\n_1\dots\m_n\n_n}(x, x_1,\dots, x_n) 
 + \frac{1}{d} \, \left(\pa\cdot K\right)\,\eta_{\m\n} \,\cS_{n+1}^{\m\n\m_1\n_1\dots\m_n\n_n}(x, x_1,\dots, x_n)
 \label{CWIgen}
\eea
where (\ref{CKeq}) has been used, in order to derive the CWIs for the $n$-point functions of the stress tensor. If this identity is 
integrated over $x$, the left hand side gives a surface term which vanishes, provided the correlation functions fall off fast enough 
at infinity, an assumption that may be checked {\it a posteriori}. On the right hand side one may use the previously derived
conservation and anomalous trace Ward Identities for the $(n+1)$-point function $\cS_{n+1}$, which are given in terms 
of the $n$-point function $\cS_{n}$ by relations of the form (\ref{twoptcons})-(\ref{threeptcons}) and (\ref{twoptr})-(\ref{threeptr}). 
By inserting the various CKV solutions of (\ref{CKeq}) into these relations, all the CWIs may be derived for arbitrary
$n$-point functions of $T^{\mu\nu}$ in the covariant geometric approach.

The Dilation CWIs correspond to the one Dilational Conformal Killing vector
\be
K_{\m}^{(D)}(x) \equiv  x_\m\,,\qquad \pa\cdot K^{(D)} = d
\ee
whereas the Special CWIs correspond to the $d$ Special Conformal Killing vectors in flat space
\be
K^{(C)\,\ka}_{\m}(x) \equiv K^{(C)\, \ka}_{\ \ \ \m}(x) \equiv 2x^{\ka}x_{\m} - x^2 \del^{\ka}_{\ \,\m}\,,
\qquad \pa\cdot K^{(C)\,\ka}(x) = (2d)\,x^{\ka}\,,\quad\ka =1, \dots, d 
\ee
each of which satisfy (\ref{CKeq}), each of which give CWIs when substituted into (\ref{CWIgen}).

\subsection{Momentum space}

When these relations above are expressed in momentum space by Fourier transforming, the results are the 
CWIs for the reduced $n$-point correlation functions defined in (\ref{Fourtrans}). The resulting general form of the CWIs in momentum space is \cite{Coriano:2012wp,Bzowski:2013sza}
\be
\left[\sum_{j=1}^n w_j  -\big(n-1\big) d -\sum_{j=1}^{n-1}p_j \cdot \frac{\pa}{\pa p_j}\right]
\tilde \cS_n^{\m_1\n_1\dots\m_n\n_n} (p_1, \dots, p_n)= X_n^{\m_1\n_1\dots\m_n\n_n} (p_1, \dots, p_n)
\label{DCWI}
\ee
and
\begin{align}
&\sum_{j=1}^{n-1}\left[2\,\big(w_j- d\big)\, \frac{\pa}{\pa p_j^\ka} -2\, p_j^\a\frac{\pa^2}{\pa p_j^\a\pa p_j^\ka}
+  p_{j\ka} \frac{\pa^2}{\pa p_j^\a\pa p_{j\a}}\right] \tilde \cS_n^{\m_1\n_1\dots\m_n\n_n} (p_1, \dots, p_n)\notag\\
& + 2\, \sum_{j=1}^{n-1}\left[\d^{\m_j}_{\ \ \ka}\frac{\pa}{\pa p_j^{\a_j}}-\eta_{\ka\a_j}\frac{\pa}{\pa p_{j\,\m_j}}\right]
\tilde\cS_n^{\m_1\n_1\dots\a_j\n_j\dots \m_n\n_n}(p_1, \dots, p_n)\notag\\
& + 2\, \sum_{j=1}^{n-1}\left[\d^{\n_j}_{\ \ \ka}\frac{\pa}{\pa p_j^{\b_j}}-\eta_{\ka\b_j}\frac{\pa}{\pa p_{j\,\n_j}}\right]
\tilde\cS_n^{\m_1\n_1\dots\m_j\b_j\dots \m_n\n_n}(p_1, \dots, p_n)
= Y_{n\,\ka}^{\m_1\n_1\dots\m_n\n_n} (p_1, \dots, p_n)
\label{SCWI}
\end{align}
where $(p_1 + \dots + p_n)^\m = 0$, $w_j=d$ is the conformal weight of $T^{\m\n}$ in $d$ dimensions, and 
$X_n^{\m_1\n_1\dots\m_n\n_n}$ and $Y_{n\,\ka}^{\m_1\n_1\dots\m_n\n_n}$ are terms involving lower $n$-point functions, 
obtained by use of the conservation and trace Ward Identities, such as (\ref{twoptcons})-(\ref{threeptcons}) and
(\ref{twoptr})-(\ref{threeptr}). In precisely $d=4$ dimensions, $w_j= 4$, and (\ref{DCWI}) and (\ref{SCWI}) take the explicit forms
\bes
\bea
&&\hspace{2cm}\left[4 -p \cdot \frac{\pa}{\pa p}\right] \cS_2^{\m_1\n_1\m_2\n_2} (p, -p) = 4\, \tilde \cA_2^{\m_1\n_1\m_2\n_2} (p, -p)\\
&&\left[ -2\, p_{\a}  \frac{\pa^2}{\pa p_{\a} \pa p_{\ka}} + p^{\ka} \frac{\pa^2}{\pa p_{\a} \pa p^\a} \right]
\tilde \cS_2^{\m_1\n_1\m_2\n_2} (p, -p)+\,  2 \left(\h^{\ka\m_1} \frac{\pa}{\pa p^{\a_1}} - \d^{\ka}_{\ \a_1} \frac{\pa}{\pa p_{\m_1}}\right) \tilde \cS_2^{\a_1\n_1\m_2\n_2} (p, -p)\nn
&& +\,  2 \left(\h^{\ka\n_1} \frac{\pa}{\pa p^{\b_1}} - \d^{\ka}_{\ \b_1} \frac{\pa}{\pa p_{\n_1}}\right) \tilde \cS_2^{\m_1\b_1\m_2\n_2} (p,-p)
 = -8\,\frac{\pa}{\pa p_{2\ka}} \tilde \cA_2^{\m_1\n_1\m_2\n_2}(p,p_2)\Big\vert_{p_2 = -p}
\eea
\label{CWIs2}
\ees
for $n=2$, while
\begin{equation}
\left[4 - p_1\cdot\sdfrac{\pa}{\pa p_1} - p_2\cdot\sdfrac{\pa}{\pa p_2} \right] \tilde\cS_3^{\m_1\n_1\m_2\n_2\m_3\n_3} (p_1, p_2, -p_1 - p_2)= 8 \, \tilde \cA_3^{\m_1\n_1\m_2\n_2\m_3\n_3}(p_1,p_2,-p_1-p_2)\label{CWIs3a}
\end{equation}
is the anomalous Dilatation CWI, and
\bea
&&\hspace{-1cm} \sum_{j=1}^2 \left[ -2\, p_{j \a}  \frac{\pa^2}{\pa p_{j\a} \pa p_{j  \ka}} + p_j^{\ka} \frac{\pa^2}{\pa p_{j\a} \pa p_j^\a} \right]
\tilde \cS_3^{\m_1\n_1\m_2\n_2\m_3\n_3} (p_1, p_2, -p_1-p_2)\nn
&& +\,  2 \left(\h^{\ka\m_1} \frac{\pa}{\pa p_1^{\a_1}} - \d^{\ka}_{\ \a_1} \frac{\pa}{\pa p_{1\m_1}}\right) \tilde \cS_3^{\a_1\n_1\m_2\n_2\m_3\n_3} (p_1, p_2, -p_1-p_2)\nn
&& +\,  2 \left(\h^{\ka\n_1} \frac{\pa}{\pa p_1^{\b_1}} - \d^{\ka}_{\ \b_1} \frac{\pa}{\pa p_{1\n_1}}\right) \tilde \cS_3^{\m_1\b_1\m_2\n_2\m_3\n_3} (p_1, p_2, -p_1-p_2)\nn
&& +\,  2 \left(\h^{\ka\m_2} \frac{\pa}{\pa p_2^{\a_2}} - \d^{\ka}_{\ \a_2} \frac{\pa}{\pa p_{2\m_2}}\right) \tilde \cS_3^{\m_1\n_1\a_2\n_2\m_3\n_3} (p_1, p_2, -p_1-p_2)\nn
&& +\,  2 \left(\h^{\ka\n_2} \frac{\pa}{\pa p_2^{\b_2}} - \d^{\ka}_{\ \n_2} \frac{\pa}{\pa p_{2\n_2}}\right) \tilde \cS_3^{\m_1\n_1\m_2\b_2\m_3\n_3} (p_1, p_2, -p_1-p_2)\nn
&&\hspace{1cm} = -16\ \frac{\pa}{\pa p_{3\ka}} \tilde \cA_3^{\m_1\n_1\m_2\n_2\m_3\n_3}(p_1,p_2,p_3)\Big\vert_{p_3 = -p_1 -p_2}\label{CWIs3b}
\eea
is the anomalous Special CWI satisfied by the $n=3$ vertex function. 

The explicit anomalous terms on the right sides of \eqref{CWIs3a} and \eqref{CWIs3b} are new results, the full
derivations of which are straightforward but somewhat lengthy and will be presented elsewhere, as they are not needed for the main analysis of the trace identities and conformal anomaly contribution to $\braket{TTT}$ in this paper. We note only that all these CWIs are finite directly in $d = 4$ dimensions and the right sides of (\ref{CWIs3a})-(\ref{CWIs3b}) are completely determined by the conservation and trace identities in the covariant approach, including all contact terms, and that in deriving \eqref{CWIs3a} and \eqref{CWIs3b} the symmetry properties of $\tilde \cS_3$ such as
\be
\tilde \cS_3^{\m_2\n_2\m_3\n_3\m_1\n_1} (p_2, p_3,p_1)  = \tilde \cS_3^{\m_1\n_1\m_2\n_2\m_3\n_3} (p_1, p_2, p_3)=  \tilde \cS_3^{\n_1\m_1\m_2\n_2\m_3\n_3} (p_1, p_2, p_3) 
\ee
have been used freely.

\section{Tensor decomposition and reconstruction of $\lag TTT\rag$}
\label{Sec:Reconst}

In this section we briefly summarize the approach of \cite{Bzowski:2013sza}, who work in general $d$ dimensions in the momentum 
representation and project the Fourier transformed CWIs (\ref{DCWI})-(\ref{SCWI}) onto their transverse, traceless parts only, assuming no 
anomalous contributions. In this approach the inhomogeneous $X_n$ and $Y_n$ terms of the CWIs may be neglected since the terms on the 
right sides of (\ref{DCWI}) and (\ref{SCWI}) involve only longitudinal and trace parts, which drop out of the transverse, traceless projection. The 
number of transverse, traceless tensors with independent form factors is also much fewer than the general $\cS_n$ (being only five for $n=3$). 
The CWIs for these $5$ scalar form factors may be solved explicitly in the momentum representation in general $d\neq 4$ dimensions in terms 
of `triple-$K$' integrals, which however contain $(d-4)^{-1}$ pole singularities, so that counterterms are needed to obtain a finite $d \rightarrow 4$ limit. 
After their subtraction the anomalous trace (\ref{Adef}) results. This amounts to rederiving the conformal anomaly (\ref{trace}) once again. Finally the
full solution for $\lag T^{\m_1\n_1} T^{\m_2\n_2}T^{\m_3\n_3}(x_3)\rag_g$ can be reconstructed in principle by use of the conservation and trace 
Ward Identities and the limit $d \rightarrow 4$ taken.

In order to apply this method to tensor operators, it is necessary to introduce the projection operators
\vspace{-5mm}
\bes
\bea
\pi^\m_{\ \,\n}(p) &\equiv &\d^\m_{\ \, \n} - \frac{p^\m p_\n}{p^2}\\
\Pi^{\m\n}_{\ \ \ \a\b}(p) &\equiv& \pi^{(\m}_{\ \ \a}(p)\,\pi^{\n)}\,_{\!\!\!\b}(p) - \frac{1}{d-1}\, \pi^{\m\n}(p)\,\pi_{\a\b}(p)
\eea\label{pidef}
\ees
onto transverse vectors and transverse, traceless tensors respectively, in $d$ dimensions. One defines further the longitudinal
and trace projectors
\bes
\bea
\La^{\m\n}_{\ \ \ \a\b}(p)&\equiv& \frac{1}{p^2} \left\{ p^{(\m} \d^{\n)}_{\ \ \a}\, p_{\b} + p^{(\m} \d^{\n)}\,_{\!\!\!\b}\,p_{\a}
- \frac{p_\a p_\b}{d-1} \left(\eta^{\m\n} + (d-2)\, \frac{p^\m p^\n}{p^2} \right)\right\} \\
\Th^{\m\n}_{\ \ \ \a\b}(p) &\equiv& \frac{\pi^{\m\n}(p)}{d-1}\,\eta_{\a\b}\label{Thdef}
\eea
\ees
such that $\Pi + \La + \Th = \id$ is the identity. Thus given any symmetric second rank tensor $T^{\m\n}(p)$, 
one may decompose it as
\be
T^{\m\n}(p) =  t^{\m\n}(p) + \La^{\m\n}(p) + \Th^{\m\n}(p)
\label{decomp}
\ee
where the first term is its transverse, traceless part
\be
t^{\m\n}(p) \equiv \Pi^{\m\n}_{\ \ \ \a\b}(p) \, T^{\a\b}(p)
\ee
while the $\La^{\m\n} \equiv \La^{\m\n}_{\ \ \ \a\b}T^{\a\b}$ and $\Th^{\m\n}\equiv \Th^{\m\n}_{\ \ \ \a\b}T^{ab}$ terms in (\ref{decomp}) 
depend only upon its longitudinal and trace contractions $p_\b T^{\a\b}$ and $\eta_{\a\b}T^{\a\b}$ respectively. The latter terms are called
`semi-local' and denoted $t^{\m\n}_{\ \ loc}$ by the authors of \cite{Bzowski:2013sza}, so that 
$T^{\m\n} = t^{\m\n} + \La^{\m\n} + \Th^{\m\n} = t^{\m\n} + t^{\m\n}_{\ \ loc}$.

The transverse, tracefree projected corrrelation functions
\be
^{(\Pi)}\tilde \cS_n^{\m_1\n_1\dots \m_n\n_n}(p_1, \dots, p_n) \equiv 
\Pi^{\m_1\n_1}_{\ \ \quad\a_1\b_1}(p_1)\dots \Pi^{\m_n\n_n}_{\ \ \quad\a_n\b_n}(p_n)\,
\tilde \cS_n^{\a_1\b_1\dots \a_n\b_n}(p_1, \dots, p_n)
\ee
obey {\it sourcefree} homogeneous Special CWIs in $d \neq 4$ dimensions with no $X_n$ or $Y_n$ terms 
on the right side of (\ref{SCWI})\cite{Bzowski:2013sza}. In order to reconstruct the full three-point correlator 
use is then made of the basic identity
\bea
T_2T_3 &=& t_2t_3 + (\La + \Th)_2 t_3 + t_2(\La + \Th)_3 + (\La + \Th)_2  (\La + \Th)_3 \nn
&=& t_2t_3 + (\La + \Th)_2 T_3 + T_2(\La + \Th)_3  - (\La + \Th)_2  (\La + \Th)_3 
\eea
following from (\ref{decomp}) for the product of two full stress tensors, suppressing indices in a symbolic notation. 
Applying this identity again, it follows that the product of three full stress tensors can be expressed
\bea
&&T_1T_2T_3 = t_1t_2t_3 + (\La + \Th)_1 T_2T_3  + T_1 (\La + \Th)_2 T_3 + T_1 T_2 (\La + \Th)_3\nn
&&\hspace{5mm} - T_1 (\La + \Th)_2(\La + \Th)_3
- (\La + \Th)_1 T_2 (\La + \Th)_3 - (\La + \Th)_1(\La + \Th)_2T_3 \nn
&&\hspace{2cm} +  (\La + \Th)_1 (\La + \Th)_2(\La + \Th)_3
\eea
or more explicitly, the full three-point correlator in momentum space is given by
\bea
&&\hspace{-9mm}\tilde \cS_3^{\m_1\n_1\m_2\n_2\m_3\n_3} = \,
^{(\Pi)}\tilde \cS_3^{\m_1\n_1\m_2\n_2\m_3\n_3} 
+ (\La + \Th)^{\m_1\n_1}_{\ \ \quad\a_1\b_1}(p_1)\,\tilde \cS_3^{\a_1\b_1\m_2\n_2\m_3\n_3}\nn
&&+\, (\La + \Th)^{\m_2\n_2}_{\ \ \quad\a_2\b_2}(p_2)\,\tilde \cS_3^{\m_1\n_1\a_2\b_2\m_3\n_3}
+ (\La + \Th)^{\m_3\n_3}_{\ \ \quad\a_3\b_3}(p_3)\,\tilde \cS_3^{\m_1\n_1\m_2\n_2\a_3\b_3}\nn
&&\hspace{1cm} -\,(\La + \Th)^{\m_2\n_2}_{\ \ \quad\a_2\b_2}(p_2)\,(\La + \Th)^{\m_3\n_3}_{\ \ \quad\a_3\b_3}(p_3)\,
\tilde \cS_3^{\m_1\n_1\a_2\b_2\a_3\b_3}\nn
&&\hspace{1cm} -\,(\La + \Th)^{\m_1\n_1}_{\ \ \quad\a_1\b_1}(p_1)\,(\La + \Th)^{\m_3\n_3}_{\ \ \quad\a_3\b_3}(p_3)\,
\tilde \cS_3^{\a_1\b_1\m_2\n_2\a_3\b_3}\nn
&&\hspace{1cm} -\, (\La + \Th)^{\m_1\n_1}_{\ \ \quad\a_1\b_1}(p_1)\,(\La + \Th)^{\m_2\n_2}_{\ \ \quad\a_2\b_2}(p_2)\,
\tilde \cS_3^{\a_1\b_1\a_2\b_2\m_3\n_3}\nn
&&+\, (\La + \Th)^{\m_1\n_1}_{\ \ \quad\a_1\b_1}(p_1)\,(\La + \Th)^{\m_2\n_2}_{\ \ \quad\a_2\b_2}(p_2)\,(\La + \Th)^{\m_3\n_3}_{\ \ \quad\a_3\b_3}(p_3)\, 
\tilde \cS_3^{\a_1\b_1\a_2\b_2\a_3\b_3}
\label{S3reconst}
\eea
where we write out the indices but continue to suppress the dependence on the momenta $(p_1,p_2,p_3)$ for brevity. When all the 
$\La + \Th$ longitudinal  and trace terms are expanded out, one gets from the $26$ terms above (excluding $^{(\Pi)}\tilde \cS_3$ itself), 
finally $7$ terms containing only the trace over one, two or three pair of indices, {\it viz.}
\bea 
&&\hspace{-1.4cm}^{(\Th)\!}\tilde \cS_3^{\m_1\n_1\m_2\n_2\m_3\n_3} \equiv 
\Th^{\m_1\n_1}_{\ \quad\a_1\b_1\!}(p_1)\,\tilde \cS_3^{\a_1\b_1\m_2\n_2\m_3\n_3}
+\, \Th^{\m_2\n_2}_{\ \quad\a_2\b_2\!}(p_2)\,\tilde \cS_3^{\m_1\n_1\a_2\b_2\m_3\n_3}\hspace{-2cm}\nn
&&+\, \Th^{\m_3\n_3}_{\ \quad\a_3\b_3\!}(p_3)\,\tilde \cS_3^{\m_1\n_1\m_2\n_2\a_3\b_3}
-\Th^{\m_2\n_2}_{\ \quad\a_2\b_2}(p_2)\,\Th^{\m_3\n_3}_{\ \ \quad\a_3\b_3\!}(p_3)\,
\tilde \cS_3^{\m_1\n_1\a_2\b_2\a_3\b_3}\nn
&&\hspace{-1.4cm} -\Th^{\m_1\n_1}_{\ \quad\a_1\b_1}(p_1)\,\Th^{\m_3\n_3}_{\ \quad\a_3\b_3\!}(p_3)\,
\tilde \cS_3^{\a_1\b_1\m_2\n_2\a_3\b_3\!}
 -\Th^{\m_1\n_1}_{\ \quad\a_1\b_1}(p_1)\,\Th^{\m_2\n_2}_{\ \quad\a_2\b_2\!}(p_2)\,
\tilde \cS_3^{\a_1\b_1\a_2\b_2\m_3\n_3}\hspace{-1cm}\nn
&&+\, \Th^{\m_1\n_1}_{\ \quad\a_1\b_1\!}(p_1)\,\Th^{\m_2\n_2}_{\ \quad\a_2\b_2\!}(p_2)\,\Th^{\m_3\n_3}_{\ \quad\a_3\b_3\!}(p_3)\, 
\tilde \cS_3^{\a_1\b_1\a_2\b_2\a_3\b_3}\,,
\label{STh}
\eea
an analogous $7$ terms with longitudinal parts only, $\La$ replacing $\Th$ everywhere in the expression (\ref{STh}) above,
and $12$ mixed terms involving both the $\La$ and $\Th$ projectors in various combinations. 

The $7$ longitudinal terms involving only $\La$ may be obtained from the conservation Ward Identity because they involve 
contractions with one or more of the $p_j$, and are thus given in terms of the two-point function by (\ref{threeptmom}), which
does not involve any traces of $\tilde \cS_3$. The $12$ mixed terms involve both at least one trace and one longitudinal projector. 
Thus from (\ref{threeptr}) a typical mixed term involves
\bea
&&\hspace{-1.5cm}\eta_{\a_1\b_1}(p_2)_{\m_2}\tilde \cS_3^{\a_1\b_1\m_2\n_2\m_3\n_3}(p_1, p_2 ,p_3) = \nn
&&\hspace{-1.2cm}- 2\,(p_2)_{\m_2}\,\tilde\cS_2^{\m_2\n_2\m_3\n_3}(p_1 +p_2,p_3) -2\,(p_2)_{\m_2}\, \tilde\cS_2^{\m_2\n_2\m_3\n_3}(p_2,p_1 + p_3) 
 + \, 8 \,(p_2)_{\m_2}\tilde \cA_2^{\m_2\n_2\m_3\n_3} (p_1).
\label{mixedLaTh}
\eea
Since as we show below in Sec.\,\ref{Sec:AnomTTT}, the second variation of the anomaly $\tilde \cA_2$ is transverse, the last term in (\ref{mixedLaTh})
vanishes upon contraction with $(p_2)_{\m_2}$. For this reason all $12$ mixed longitudinal/trace terms in the expansion of (\ref{S3reconst}) are also independent 
of the second trace anomaly variation $\tilde \cA_2$, and determined solely by the conservation Ward Identities plus the trace identities in terms of the two-point 
function $\tilde \cS_2$. This leaves only the $7$ trace terms in (\ref{STh}) which do depend upon the trace anomaly second variation $\tilde \cA_2$, upon which 
we focus here.

Substituting the definition of $\Th$ projector (\ref{Thdef}) for $d=4$ enables us to write the $7$ terms in (\ref{STh}) in the form
\bea 
&&\hspace{-1.4cm}^{(\Th)\!}\tilde \cS_3^{\m_1\n_1\m_2\n_2\m_3\n_3} = 
\sdfrac{1}{3}\, \pi^{\m_1\n_1}(p_1)\, \eta_{\a_1\b_1}\,\tilde \cS_3^{\a_1\b_1\m_2\n_2\m_3\n_3}
+\, \sdfrac{1}{3}\, \pi^{\m_2\n_2}(p_2)\, \eta_{\a_2\b_2}\,\tilde \cS_3^{\m_1\n_1\a_2\b_2\m_3\n_3}\hspace{-2cm}\nn
&&+\, \sdfrac{1}{3}\, \pi^{\m_3\n_3}(p_3)\, \eta_{\a_3\b_3}\,\tilde \cS_3^{\m_1\n_1\m_2\n_2\a_3\b_3}
- \sdfrac{1}{9}\, \pi^{\m_2\n_2}(p_2) \pi^{\m_3\n_3}(p_3)\,\eta_{\a_2\b_2}\, \eta_{\a_3\b_3}\,
\tilde \cS_3^{\m_1\n_1\a_2\b_2\a_3\b_3}\nn
&&\hspace{-1.4cm} -\sdfrac{1}{9}\, \pi^{\m_1\n_1}(p_1)\,\pi^{\m_3\n_3}(p_3)\, \eta_{\a_1\b_1}\,\eta_{\a_3\b_3}\,
\tilde \cS_3^{\a_1\b_1\m_2\n_2\a_3\b_3\!}
 -\sdfrac{1}{9}\, \pi^{\m_1\n_1}(p_1) \, \pi^{\m_2\n_2}(p_2)\,\eta_{\a_1\b_1}\,\eta_{\a_2\b_2}\,
\tilde \cS_3^{\a_1\b_1\a_2\b_2\m_3\n_3}\hspace{-1cm}\nn
&&+\, \sdfrac{1}{27} \,\pi^{\m_1\n_1}(p_1)\,  \pi^{\m_2\n_2}(p_2)\,\pi^{\m_3\n_3}(p_3)\, \eta_{\a_1\b_1}\, \eta_{\a_2\b_2}\,\eta_{\a_3\b_3}
\tilde \cS_3^{\a_1\b_1\a_2\b_2\a_3\b_3}
\label{STrace}
\eea
containing one, two and three trace terms. Rather than explicitly adding counterterms in the dimensional regularization procedure
of \cite{Bzowski:2013sza}, in effect rederiving the conformal anomaly, we may more directly make use of the trace Ward Identity 
(\ref{threeptr}) for the three-point function derived from the known form of the local anomaly (\ref{trace}). From this one
observes that (\ref{STrace}) contains terms that may be expressed in terms of the two-point function $\tilde \cS_2$, and
which for present purposes may be considered non-anomalous, and in addition, the last term of (\ref{threeptr}), with
an explicit dependence upon the second variation of the anomaly $\tilde \cA_2$. Separating out these terms in (\ref{STrace})
explicitly dependent upon $\tilde \cA_2$ we have 
\bea 
&&\hspace{-1.4cm}^{(\Th)\!}\tilde \cS_3^{\m_1\n_1\m_2\n_2\m_3\n_3}(p_1,p_2,p_3)\Big\vert_{\cA_2} = 
\sdfrac{4}{3}\, \pi^{\m_1\n_1}(p_1)\, \tilde \cA_2^{\m_2\n_2\m_3\n_3}(p_2,p_3)
+\, \sdfrac{4}{3}\, \pi^{\m_2\n_2}(p_2)\,\tilde \cA_2^{\m_1\n_1\m_3\n_3}(p_1,p_3)\hspace{-2cm}\nn
&&+\, \sdfrac{4}{3}\, \pi^{\m_3\n_3}(p_3)\, \tilde \cA_2^{\m_1\n_1\m_2\n_2}(p_1,p_2)
- \sdfrac{4}{9}\, \pi^{\m_2\n_2}(p_2) \pi^{\m_3\n_3}(p_3)\,\eta_{\a_3\b_3}\,
\tilde \cA_2^{\m_1\n_1\a_3\b_3}(p_1,p_3)\nn
&&\hspace{-1.4cm} -\sdfrac{4}{9}\, \pi^{\m_1\n_1}(p_1)\,\pi^{\m_3\n_3}(p_3)\, \eta_{\a_1\b_1}\,
\tilde \cA_2^{\a_1\b_1\m_2\n_2\!}(p_1,p_2)
 -\sdfrac{4}{9}\, \pi^{\m_1\n_1}(p_1) \, \pi^{\m_2\n_2}(p_2)\,\eta_{\a_2\b_2}\,
\tilde \cA_2^{\a_2\b_2\m_3\n_3}(p_2,p_3)\hspace{-1cm}\nn
&&+\, \sdfrac{4}{27} \,\pi^{\m_1\n_1}(p_1)\,  \pi^{\m_2\n_2}(p_2)\,\pi^{\m_3\n_3}(p_3)\, \eta_{\a_1\b_1}\, \eta_{\a_2\b_2}\,
\tilde \cA_2^{\a_1\b_1\a_2\b_2}(p_1,p_2)
\label{STraceA2}
\eea
which we shall show is identical to the contribution to $\tilde \cS_3$ determined from the non-local anomaly effective action of
\cite{Riegert:1987kt,Antoniadis:1992xu} directly.

\section{Anomaly action and total effective action for CFTs}
\label{Sec:AnomAct}

The general form of the anomalous trace (\ref{trace}) is a consequence of locality of the underlying QFT,  and the Wess-Zumino consistency condition, 
which amounts to the requirement that a covariant action functional $\cS_{\rm anom}[g]$ of the full metric $g_{\m\n}$ must exist such that
\be
\cS_{\rm anom}[e^{2\s}\bar g] = \cS_{\rm anom}[\bar g] + \G_{WZ}[\bar g;\s]
\label{WZGamma}
\ee
for arbitrary $g_{\m\n}(x) = e^{2\s (x)} \,\bar g_{\m\n}(x)$, and whose conformal variation
\bea
&&2\, g_{\m\n}(x)\,\frac{\del \cS_{\rm anom}[g]}{\del g_{\m\n}(x)} =  \frac{\del \cS_{\rm anom}[e^{2\s}\bar g]}{\del \s (x)} \Big\vert_{g = e^{2\s}\bar g}
=  \frac{\del \G_{WZ}[\bar g;\s]}{\del \s (x)}\nn
&& \hspace{1.5cm}=\cA \equiv \sqrt{-g}\, \Big\{ b\, C^2 + b' \big(E - \tfrac{2}{3}\sq R\big) \Big\}\Big\vert_{g = e^{2\s}\bar g}
\label{traceiden}
\eea
is the anomaly. The admixture of the $\sq R$ term with specific coefficient $-\frac{2}{3}$ in (\ref{traceiden}) is chosen so that both invariants 
in (\ref{traceiden}) have simple dependences upon $\s$
\bes
\bea
\sqrt{-g}\,C^2 &=& \sqrt{-\overline g}\,\overline C^2\,\label{Csig}\\
\sqrt{-g}\,\left(E - \tfrac{2}{3}\sq R\right) &=& \sqrt{-\overline g}\,
\left(\overline E - \tfrac{2}{3}\sqb\bar R\right) + 4\,\sqrt{-\overline g}\, \bar\D_4\,\s  \label{Esig}
\eea
\label{CEsig}
\ees
in the local conformal parametrization $g_{\m\n} = e^{2\s} \bar g_{\m\n}$.

The fourth order differential operator 
\be
\D_4 \equiv \na_\m \left(\na^\m\na^\n +2R^{\m\n} - \tfrac{2}{3} R g^{\m\n} \right)\na_\n
=\sq^2 + 2 R^{\m\n}\na_\m\na_\n - \tfrac{2}{3} R \sq + \tfrac{1}{3} (\na^\m R)\na_\m
\label{Deldef}
\ee
is the unique fourth order scalar kinetic operator that is conformally covariant
\be
\sqrt{-g}\, \D_4 = \sqrt{-\bar g}\, \bar \D_4
\label{invfour}
\ee
for arbitrary $\s(x)$ \cite{Riegert:1987kt,Antoniadis:1992xu,Antoniadis:1991fa,Mazur:2001aa,Panietz}. Because of the simple dependences 
(\ref{CEsig}), the Wess-Zumino functional in (\ref{WZGamma}) quadratic in $\s$ is easily found to be
\bea
&&\G_{_{\!W\!Z}}[\bar g;\s] = 2 b'\! \int\,d^4x\,\sqrt{-\bar g}\ \s\,\bar\D_4\,\s + \int\,dx\ \overline{\!\cA} \, \s\nn
&&\quad = b' \int\,d^4x\,\sqrt{-\bar g}\,\Big[2\,\s\,\bar\Delta_4\,\s + \left(\bar E - \tfrac{2}{3} \sqb \bar R\right)\s \Big]
+ b \int\,d^4x\,\sqrt{-\bar g}\, \bar C^2\,\s
\label{WZfour}
\eea
by inspection, up to terms which are $\s$ independent, {\it i.e.} conformally invariant, and hence do not contribute
to the variation in (\ref{traceiden}). Moreover, solving (\ref{Esig}) for $\s$ by inverting the differential operator (\ref{Deldef})
in favor of its Green's function $D_4(x,x') = (\D_4^{-1})_{xx'}$ we find
\be
\G_{_{\!W\!Z}}[\bar g;\s] = \cS_{\rm anom}^{^{NL}}[g=e^{2\s}\bar g] - \cS_{\rm anom}^{^{NL}}[\bar g],
\label{Weylshift}
\ee
with
\be
\cS_{\rm anom}^{^{NL}}[g] =\sdfrac {1}{4}\!\int \!d^4x\sqrt{-g_x}\, \Big(E - \tfrac{2}{3}\sq R\Big)_{\!x} 
\int\! d^4x'\sqrt{-g_{x'}}\,D_4(x,x')\bigg[\sdfrac{b'}{2}\, \big(E - \tfrac{2}{3}\sq R\big) +  b\,C^2\bigg]_{x'}
\label{Snonl}
\ee
which is a non-local form of the exact quantum 1PI effective action of the anomaly. This non-locality in terms only of curvature invariants 
and arising from the Green's function inverse of $\D_4$ cannot be removed by any addition of local terms to the anomaly action, 
such as the $\int R^2$ associated with $b''$ term in (\ref{trace}) by (\ref{varRsq}).

 It is possible to add to $\cS_{\rm anom}^{^{NL}}$ 
arbitrary Weyl invariant terms (local or not) which drop out of the difference in (\ref{Weylshift}), but these also cannot remove the 
non-locality in the essential Weyl non-invariant part of the anomaly action (\ref{Snonl}). In particular, if the non-local term Weyl invariant term
\be 
\sdfrac {b^2}{8b'}\!\int \!d^4x\sqrt{-g_x}\, \big(C^2\big)_{x} \int\! d^4x'\sqrt{-g_{x'}}\,D_4(x,x') \big(C^2\big)_{x'}
\label{SaddW}
\ee
is added to (\ref{Snonl}) to complete the square, we obtain
\bea
&&\hspace{-8mm}\cS_{\rm anom}^{^{NL}}[g] \rightarrow \sdfrac {b'}{8\,}\!\int \!d^4x\,\sqrt{-g_x}\int\! d^4x'\,\sqrt{-g_{x'}}\, 
\bigg[\big(E - \tfrac{2}{3}\sq R\big) +  \sdfrac{b}{\,b'}\,C^2\bigg]_{x}\!
D_4(x,x')\bigg[\big(E - \tfrac{2}{3}\sq R\big) +  \sdfrac{b}{\,b'}\,C^2\bigg]_{x'}\nn
&&\hspace{2cm} = \sdfrac {1}{8b'}\int d^4x \int d^4x'\, \cA(x)\, D_4(x,x')\, \cA(x')
\label{Snonlsq}
\eea
and it becomes possible to recast the generally covariant non-local effective action (\ref{Snonlsq}) in local form by the introduction of  
only a single new scalar field $\vf$, called the {\it conformalon} field \cite{Mottola:2016mpl}. Because the minimal non-local action 
(\ref{Snonl}) without the addition Weyl invariant term (\ref{SaddW}) is asymmetrical in the invariants $E$ and $C^2$, two
scalar fields would be necessary to render (\ref{Snonl}) into a local form \cite{Mottola:2006ew,Shapiro:1994ww}. Since the 
anomalous effective action is determined only up to such Weyl invariant terms in any case, adding the Weyl invariant 
term (\ref{SaddW}) does not affect the trace anomaly or otherwise affect our conclusions on the trace terms in $\lag TTT\rag$. 

Thus, if the anomaly effective action is recast in local form
\bea
&&\hspace{-1.5cm} \cS_{\rm anom}[g;\vf] \equiv -\sdfrac{b'}{2} \int d^4x\,\sqrt{-g}\, \Big[ (\sq \vf)^2 - 2 \big(R^{\m\n} - \tfrac{1}{3} R g^{\m\n}\big)
(\na_\m\vf)(\na_\n \vf)\Big]\nn
&& \hspace{1.5cm} +\, \sdfrac{1}{2}\,\int d^4x\,\sqrt{-g}\  \Big[b'\big(E - \tfrac{2}{3}\sq R\big) + b\,C^2 \Big]\,\vf
\label{Sanom}
\eea
by the introduction of a scalar conformalon $\varphi$, and varied with respect to $\varphi$, the linear eq. of motion
\be
\sqrt{-g}\,\D_4\, \vf = \sqrt{-g}\left[\sdfrac{E}{2}- \sdfrac{\!\sq R\!}{3} + \sdfrac{b}{2b'}\, C^2\right] = \frac{1}{2b'} \,\cA
\label{phieom}
\ee
results. Multiplication by $\sqrt{-g}$ in (\ref{phieom}) is convenient because of the property (\ref{invfour}). If \eqref{phieom} is solved for $\vf$ by formally inverting $\D_4$ and substituting the result into (\ref{Sanom}) the non-local effective action $\cS^{^{NL}}_{anom}[g]$ (\ref{Snonlsq}) is reproduced up to surface terms. The non-locality of \eqref{Snonlsq} or \eqref{Snonl} is thus associated with the $\Delta_4$ propagator of the scalar conformalon, a conformal collective mode not present in the classical Einstein-Hilbert gravitational action.

It is clear that linear shifts in the scalar conformalon $\vf$ are related to conformal transformations of the spacetime metric, and indeed the Wess-Zumino consistency condition implies the non-trivial relation
\be
\cS_{\rm anom}[g; \vf + 2 \s] = \cS_{\rm anom}[e^{-2 \s} g; \vf] + \cS_{\rm anom}[g; 2 \s] 
\label{SanomWZ}
\ee
for $\cS_{\rm anom}[g; \vf]$. The identity (\ref{SanomWZ}) exposes the relationship of $\varphi$ to variations
of the conformal frame of the metric, motivates the term {\it conformalon}\,  field, and also serves to distinguish $\varphi$ from dilatons and dilaton-like fields that arise in other contexts. Note that although $\vf$ is closely related to and couples to the conformal part of the metric tensor, $\vf$ is an independent spacetime
scalar field and the local action (\ref{Sanom}) is fully coordinate invariant, unlike $\G_{_{\!W\!Z}}$ in (\ref{WZfour}) which 
depends separately upon $\bar g_{\m\n}$ and $\s$, and is therefore conformal frame dependent. Because of the fourth 
order kinetic term the scalar conformalon $\vf$ has canonical mass dimension zero, which also distinguishes
it from other dilaton-like fields. It is the local form of the anomaly action (\ref{Sanom}) in terms of the scalar 
conformalon field $\vf$ that we shall vary in order to obtain the anomalous contributions to the Ward 
Identities of a CFT directly in four dimensions.

The exact 1PI quantum effective action for a CFT includes arbitrary Weyl invariant terms, which are also generally
non-local, as well as possible local terms. Thus the full effective action can be written 
\be
\cS = \cS_{\rm local}[g] + \cS_{\rm inv}[g] + \cS_{\rm anom}[g;\vf]
\label{genS}
\ee
where $\cS_{\rm local}[g]$ contains the local term $\int R^2$ term, whose conformal variation (\ref{varRsq}) 
is associated with the $b'' \sq R$ term in (\ref{trace}) (as well as the local $\int R^2$ and $\int C^2$ counterterms 
up to dimension four needed to renormalize the Einstein-Hilbert action of classical General Relativity). The
arbitrary Weyl invariant term
\be
\cS_{\rm inv}[e^{2\s}g] = \cS_{\rm inv}[g]
\ee
an example of which is (\ref{SaddW}), is responsible for the non-anomalous CWIs (\ref{DCWI}) and (\ref{SCWI}), absent 
any anomalous trace terms, whereas $\cS_{\rm anom}[g;\vf]$ given by (\ref{Sanom}) is responsible for the anomalous trace (\ref{Adef}). 
The form (\ref{genS}) of the decomposition of the quantum effective action was arrived at in \cite{Mazur:2001aa} by consideration 
of the abelian group of local Weyl shifts, and its cohomology. The local and Weyl invariant terms are elements of the trivial cohomology 
of the local Weyl group, while (\ref{Snonl}) or $\cS_{\rm anom}[g;\vf]$ are elements of the non-trivial cocycles of this cohomology. 
There are two such cocycles in $d=4$, uniquely specified by the $b$ and $b'$ anomaly coefficients 
\cite{Bonora:83,Antoniadis:1992xu,Karakhanian:1994yd,Arakelian:1995ye,Mazur:2001aa}. 

Since each term in (\ref{genS}) is invariant under general coordinate transformations, each separately obeys
the conservation Ward Identities of Sec. \ref{Sec:ConsWI}. The local and Weyl invariant terms each separately
satisfies the Trace Ward Identities, with no anomalous $\cA$ terms. The Weyl invariant
term likewise satisfies all the Conformal Ward Identities of Sec. \ref{Sec:CWI} with no anomalous contribution. All
contributions to the Ward Identities that are anomalous can only come from $\cS_{\rm anom}$, whose anomalous contributions therefore may be considered separately from the other local and Weyl invariant terms in (\ref{genS}).

A non-trivial test of the decomposition (\ref{genS}), the reasoning leading to it, and the correctness of the anomaly 
action (\ref{Sanom}) is afforded by the reconstruction algorithm of \cite{Bzowski:2013sza} for $\lag TTT\rag$ in CFTs in flat 
spacetime, in that the anomalous trace Ward Identities obeyed by $\lag TTT\rag$, encoded in the trace dependent 
contributions to its full reconstruction $^{(\Th)\!}\tilde \cS_3$ of (\ref{S3reconst}) and \eqref{STraceA2}, must come entirely from variations of $\cS_{\rm anom}[g;\vf]$. This is verified explicitly in the next several sections.

\section{Variation of the anomaly effective action}
\label{Sec:VarAct}

In order to obtain the contributions of the anomaly effective action (\ref{Sanom}) to the three-point function
$\cS_3^{\m_1\n_1\m_2\n_2\m_3\n_3}$ we require the expansion of $\cS_{\rm anom}[g;\vf]$ to third order
in deviations from flat space. The consistent expansion  of $\cS_{\rm anom}[g;\vf]$ around flat space is 
defined by the simultaneous expansion of the metric $g_{\m\n}$ and conformalon scalar $\vf$
\bes
\bea
g_{\m\n} &=& g_{\m\n}^{(0)} + g_{\m\n}^{(1)} + g_{\m\n}^{(2)} + \dots \equiv \eta_{\m\n} + h_{\m\n} + h_{\m\n}^{(2)} + \dots\\
\vf &=& \vf^{(0)} +  \vf^{(1)} +  \vf^{(2)}  + \dots
\eea
\ees
substitution of these expansions into (\ref{phieom}) and identification of terms of the same order. Thus to the lowest three orders of the expansion we have
\bes
\bea
&&\hspace{4cm}\sqb^2 \vf^{(0)} = 0 \label{eom0}\\
&&\hspace{-1.5cm}(\sqrt{-g} \D_4)^{(1)} \vf^{(0)} + \sqb^2 \vf^{(1)} = \left[\sqrt{-g}
\left( \sdfrac{E}{2}- \sdfrac{\!\sq R\!}{3} + \sdfrac{b}{2b'}\, C^2 \right)\right]^{(1)}
= - \sdfrac{\!1\!}{3}\, \sqb R^{(1)} \label{eom1}\\
&&\hspace{-2cm}(\sqrt{-g} \D_4)^{(2)} \vf^{(0)} + (\sqrt{-g} \D_4)^{(1)} \vf^{(1)} + \sqb^2 \vf^{(2)} =
\left[\sqrt{-g}\left(\sdfrac{E}{2}- \sdfrac{\!\sq R\!}{3} + \sdfrac{b}{2b'}\, C^2 \right)\right]^{(2)} \nn
&&\hspace{2cm}= \sdfrac{1}{2}E^{(2)} - \sdfrac{1}{3}\, [\sqrt{-g}\sq R]^{(2)} + \sdfrac{b}{2b'}\, [C^2]^{(2)} \label{eom2}
\eea
\ees
where $\sqb$ is the d'Alembert wave operator in flat Minkowski spacetime, and we have used the fact that $E$ and $C^2$ are second order in curvature invariants while the Ricci scalar $R$ starts at first order. 

Clearly the eq. of motion (\ref{eom0}) possesses the trivial solution $\vf^{(0)}=0$ in flat spacetime $g_{\m\n}^{(0)} = \eta_{\m\n}$, 
which corresponds to choosing boundary conditions appropriate for the standard Minkowski space vacuum, or equivalently defining 
the quantum effective action such that its first variation and the one-point function $\lag T^{\m\n}\rag_\eta = 0$ vanishes in flat spacetime 
with no boundaries. With this condition we may immediately solve the next order eq. (\ref{eom1}) to obtain
\be
\vf^{(1)} = - \sdfrac{\!1 \!}{3\sqb} \, R^{(1)}
\label{vf1}
\ee
and hence the solution of (\ref{eom2}) is
\be
\vf^{(2)} = \sdfrac{1}{\sqb^2} \left\{ (\sqrt{-g} \D_4)^{(1)}\sdfrac{\!1 \!}{3\sqb} \, R^{(1)} 
+  \sdfrac{1}{2}E^{(2)} - \sdfrac{1}{3}\, [\sqrt{-g}\sq R]^{(2)} + \sdfrac{b}{2b'}\, [C^2]^{(2)}\right\}
\label{vf2}
\ee
the latter of which contains a coincident double $\sqb^{-2}$ pole in the momentum representation to second order 
in the expansion around flat space.

Next we substitute these solutions into the anomaly action (\ref{Sanom}), also retaining terms up to third order in the expansion.
In this way we obtain the quadratic term
\be
\cS_{\rm anom}^{(2)} = - \sdfrac{b'}{2} \,\int d^4x \, \vf^{(1)} \sqb^2 \vf^{(1)} + \sdfrac{b'}{2}\,\int d^4x \, \left( - \sdfrac{2}{3} \sqb R^{(1)}\right) \vf^{(1)}
= \sdfrac{b'}{18} \,\int d^4x \, \left(R^{(1)}\right)^2
\label{Sanom2}
\ee
which is purely local, since all propagators cancel. Thus the two-point correlation function of stress tensors will contain no
poles in momentum space due to the conformal anomaly, consistent with explicit calculations \cite{AndMolMott:2003}.

The third order terms in the expansion of the anomaly action are
\bea
&&\cS_{\rm anom}^{(3)} =  - \sdfrac{b'}{2} \int d^4x \, \left\{2\,\vf^{(1)} \sqb^2 \vf^{(2)} +\vf^{(1)} \big(\sqrt{-g} \D_4\big)^{(1)} \,\vf^{\!(1)} \right\}\nn
&&\hspace{-1cm} + \sdfrac{b'}{2} \int d^4x \left\{\left( - \sdfrac{2}{3} \sqb R^{(1)}\right) \vf^{(2)} + \left(E^{(2)} - \sdfrac{2}{3}\, \sqrt{-g}\sq R\right)^{\!(2)} \vf^{(1)} \right\}
+  \sdfrac{b}{2} \int d^4x \, (C^2)^{(2)}\,\vf^{(1)} 
\label{Sanom3a}
 \eea
from which we observe that the two terms involving $\vf^{(2)}$ cancel, upon making use of (\ref{vf1}). Thus there are no coincident $\sqb^{-2}$
propagator terms in the anomaly action to third order in the expansion. The remaining terms in (\ref{Sanom3a}) yield
\bea
&&\cS_{\rm anom}^{(3)} =- \sdfrac{b'}{18} \int d^4x \, \left\{R^{(1)}\sdfrac{1}{\sqb} \big(\sqrt{-g} \D_4\big)^{\!(1)} \,\sdfrac{1}{\sqb} R^{(1)} \right\}
- \sdfrac{b'}{6} \int d^4x \left(E- \sdfrac{2}{3}\, \sqrt{-g}\sq R\right)^{\!(2)}\,\sdfrac{1}{\sqb} R^{(1)}\nn
&&\hspace{2cm} -  \sdfrac{b}{6} \int d^4x \, [C^2]^{(2)}\,\sdfrac{1}{\sqb} R^{(1)}
\eea
the first term of which involves
\be
\big(\sqrt{-g} \D_4\big)^{\!(1)} =  \big(\sqrt{-g} \sq^2\big)^{\!(1)} + 2\, \pa_{\m} \left(R^{\m\n} - \sdfrac{1}{3} \eta^{\m\n} R\right)^{\!(1)}\pa_{\n}
\ee
where (\ref{Deldef}) has been used. With the latter term one may integrate by parts and obtain 
\bea
&&\hspace{-1.1cm}\cS_{\rm anom}^{(3)}\! =\!- \sdfrac{b'}{18}\! \int\! d^4x \left\{\!R^{(1)}\!\sdfrac{1}{\sqb} \big(\sqrt{-g} \sq^2\big)^{\!(1)}\sdfrac{1}{\sqb} R^{(1)}\! \right\}
+ \sdfrac{b'}{9}\! \int\! d^4x \left\{\!\pa_{\m} R^{(1)}\!\sdfrac{1}{\sqb} \! \left(\!R^{(1)\m\n}\! - \!\sdfrac{1}{3} \eta^{\m\n} R^{(1)}\!\right)\!
\sdfrac{1}{\sqb}\pa_{\n} R^{(1)}\!\right\}\hspace{-5mm}\nn 
&&\hspace{-8mm} - \sdfrac{1}{6}\! \int\! d^4x \left(b'\, E^{\!(2)} + b\,  [C^2]^{(2)}\right)\sdfrac{1}{\sqb}R^{(1)}
+ \sdfrac{b'}{9} \!\int\! d^4x\,  R^{(1)}  \sdfrac{1}{\sqb} \left(\sqrt{-g}\sq\right)^{\!(1)}R^{(1)}
+ \sdfrac{b'}{9} \! \int\! d^4x\, R^{\!(2)}R^{(1)}
\label{Sanom3b}
\eea
which contains only single propagator poles. Finally making use of the chain rule relation
\be
\big(\sqrt{-g} \sq^2\big)^{\!(1)} =  \left[\frac{(\sqrt{-g} \sq)^2}{\sqrt{-g}}\right]^{\!(1)} 
= 2\, \sqb \big(\sqrt{-g} \sq\big)^{\!(1)} - \big(\sqrt{-g}\big)^{(1)} \sqb^2
\ee
we find that the first and next-to-last terms of (\ref{Sanom3b}) combine and partly cancel, obtaining finally
\bea
&&\hspace{-5mm} \cS_{\rm anom}^{(3)} =
 \sdfrac{b'}{9} \int\! d^4x \int\!d^4x'\!\int\!d^4x''\!\left\{\big(\pa_{\m} R^{(1)})_x\left(\sdfrac{1}{\sqb}\right)_{\!xx'}  
 \!\left(R^{(1)\m\n}\! - \!\sdfrac{1}{3} \eta^{\m\n} R^{(1)}\right)_{x'}\!
\left(\sdfrac{1}{\sqb}\right)_{\!x'x''}\!\big(\pa_{\n} R^{(1)})_{x''}\right\}\nn
&&\hspace{-6mm}- \sdfrac{1}{6}\! \int\! d^4x\! \int\!d^4x'\! \left(b'\, E^{\!(2)} + b\,  [C^2]^{(2)}\right)_{\!x}\! \left(\sdfrac{1}{\sqb}\right)_{\!xx'} \!R^{(1)}_{x'}
 + \sdfrac{b'}{18} \! \int\! d^4x\, R^{(1)}\left(2\, R^{\!(2)} + (\sqrt{-g})^{(1)} R^{(1)}\right)
\label{S3anom3}
\eea
where the last term is purely local. In fact, this last local third order term together with the second order
term (\ref{Sanom2}) may be recognized as the expansion up to to third order of the covariant local action
\be
\sdfrac{b'}{18}  \int\! d^4x\, \sqrt{-g} \,R^2
\label{bprimelocal}
\ee
which if subtracted from $S_{\rm anom}$ in (\ref{Sanom}) and upon using \eqref{varRsq}, would cancel the $- \frac{2b'}{3} \sq R$ contribution to the conformal anomaly in \eqref{Adef} resulting from $S_{\rm anom}$, leaving just $b'\, E + b \,C^2$ for the trace. 

The result (\ref{S3anom3}) for the anomaly action expanded to third order around flat space may be derived equally well 
from the non-local form (\ref{Snonlsq}) by variation of the Green's function $D_4(x,x')$. We have used the local conformalon representation of the anomaly action \eqref{Sanom} both because of its technical advantages and to illustrate the conceptual equivalence to the non-local form. See also \cite{Mirzabekian:1995qf} for comparison to the total variation of the effective action up to third order from flat space.

We comment in passing that because the local $\int R^2$ term has the conformal $\s$ dependence
\bea
&&\int d^4x \sqrt{-g} R^2 \big\vert_{g = \bar g e^{2\s}} = \int d^4x \Big[ \bar R - 6 \sqb \s -6\, \bar g^{\m\n} \pa_\m \s \pa_\n\s\Big]^2 = \\
&&\hspace{-9mm} \!\int\!\! d^4x\sqrt{-\bar g} \Big\{\bar R^2 + 36(\sqb \s)^2 +36(\bar g^{\m\n} \pa_\m \s \pa_\n\s)^2  
- 12 \bar R \sqb \s -12 \bar R \bar g^{\m\n} \pa_\m \s \pa_\n\s 
- 72 (\sqb \s)(\bar g^{\m\n} \pa_\m \s \pa_\n\s)\Big\}\nonumber
\eea
the subtraction of the particular local term (\ref{bprimelocal}) from the Wess-Zumino action (\ref{WZfour}) gives
\bea
&& \G_{_{WZ}} [\bar g; \s]  - \sdfrac{b'}{18} \int d^4x \sqrt{-g} R^2 \big\vert_{g = \bar g e^{2\s}} =  b \int\,d^4x\,\sqrt{-\bar g}\, \bar C^2\,\s 
- \sdfrac{b'}{18} \int d^4x\,\sqrt{-\bar g} \bar R^2 \nn
&&\hspace{-1cm}+ b'\! \int\,d^4x\,\sqrt{-\bar g}\,\bigg\{\! \bar E \s + 4\,\left( R^{\m\n} - \sdfrac{1}{2} R \bar g^{\m\n}\right)(\pa_{\m}\s) (\pa_{\n}\s) - 4\,(\pa \s)^2(\sqb \s)
+ 2 \left[(\pa \s)^2\right]^2\!\bigg\} 
\label{dilaton}
\eea
which essentially reproduces (up to some algebraic errors) the form of $W_loc$ of the dilaton effective action reported in \cite{Schwimmer:2010za}, after 
identification of $\s \rightarrow - \tau$ and $\bar g \rightarrow g$, $b\rightarrow c$, $b'\rightarrow -a$, and dropping the $\int \bar R^2$ term. In this particular combination the fourth order kinetic term for $\s$ in $\G_{_{WZ}} [\bar g; \s]$ is cancelled. However the Wess-Zumino functional $\G_{_{WZ}} [\bar g; \s]$
is not a covariant functional of the {\it full} physical metric $g_{\m\n}= e^{2\s}\bar g_{\m\n}$, since it was obtained from (\ref{CEsig}) by treating 
$\s$ as a field independent of the fiducial fixed base metric $\bar g_{\m\n}$. The W-Z functional is more properly regarded as non-trivial cocycle of the cohomology of the 
local Weyl group, which is closed but not exact \cite{Bonora:83,Antoniadis:1992xu,Karakhanian:1994yd,Arakelian:1995ye,Mazur:2001aa}. 
Thus it is only a Weyl variation of some fully covariant solution of the Wess-Zumino consistency, of which (\ref{Snonl}) is one representative,
and which is necessarily non-local if expressed purely in terms of the full physical metric $g_{\m\n}$. The addition of a local term, {\it i.e.} 
a trivial element of the cohomology, with any coefficient does not change the non-local pole structure of (\ref{Snonl}), and (\ref{dilaton}) 
is not an acceptable frame invariant anomaly effective action, unless one assumes that $\s\rightarrow \tau$ is to be regarded as 
a completely new dilaton field, with its own specified transformation under local Weyl transformations, and not the conformal factor 
piece of the physical spacetime metric \cite{Schwimmer:2010za}. This is quite different than the present treatment where one straightforwardly 
evaluates the 1PI effective action of the conformal anomaly and its variations in either its non-local (\ref{Snonl}) or local representation (\ref{Sanom}),
and no assumptions about any additional dilaton field(s), their interactions, or spontaneous symmetry breaking are made.

The action functional (\ref{S3anom3}), and in particular dropping its last local term is the expression we
need in order to obtain the non-local trace anomaly pole contribution to the three-point function.
It shows that single pole terms (appearing twice, but no coincident double pole $\sqb^{-2}$ terms) are expected for the trace anomaly 
contribution $^{(\Th)}S_3^{\m_1\n_1\m_2\n_2\m_3\n_3}$ to the three-point function.

\section{The Anomaly Action Contribution to $\lag TTT\rag$}
\label{Sec:AnomTTT}

Since by (\ref{EFdef}), both $E$ and $C^2$ are second order in curvature tensors, it suffices in (\ref{S3anom3}) to compute the Riemann tensor to first order
\be
R_{\m\a\n\bet}^{(1)} = \sdfrac{1}{2}\, \Big\{\!- \pa_\a\pa_\bet h_{\m\n}- \pa_\m\pa_\n h_{\a\bet} + \pa_\a\pa_\n h_{\bet\m}
+ \pa_\bet\pa_\m h_{\a\n}\Big\}
\ee
in the metric variation $h_{\m\n}$. All contractions may then be carried out with the use of the lowest order
flat space Minkowski metric $\eta^{\m\n}$. In momentum space
\be
\int d^4x \, e^{ip\cdot x} \, R_{\m\a\n\bet}^{(1)}(x) \equiv \big[R_{\m\a\n\bet}^{(1)}\big]^{\m_1\n_1}(p)\, \tilde h_{\m_1\n_1}(p)
\label{RieFour}
\ee
serves to defines the tensor polynomial
\be
\big[R_{\m\a\n\b}^{(1)}\big]^{\m_1\n_1}(p) = \sdfrac{1}{2}\, \Big\{\d^{(\m_1}_\a\, \d^{\n_1\!)\hspace{-4pt}}\,_\b\,p_\m\, p_\n
+ \d^{(\m_1}_\m\, \d^{\n_1)}_\n\,p_\a\, p_\b  - \d_\b\,^{\hspace{-2pt}(\m_1}\, \d^{\n_1)}_\m\,p_\a \,p_\n - \d^{(\m_1}_\a\, \d^{\n_1)}_\n\,p_\b \,p_\m  \Big\}
\label{Riemom}
\ee
which has the contractions
\be
\big[R^{(1)}_{\m\n}\big]^{\m_1\n_1}(p) = \eta^{\a\b}\, \big[R_{\m\a\n\b}^{(1)}\big]^{\m_1\n_1}(p)
= \sdfrac{1}{2}\,\Big\{\eta^{\m_1\n_1}\,p_\m\, p_\n  + \d^{(\m_1}_\m\, \d^{\n_1)}_\n\,p^2 
 - p^{(\m_1}\, \d^{\n_1)}_\m\,p_\n- p^{(\m_1} \d^{\n_1)}_\n\,p_\m \Big\} 
 \label{Riccip}
\ee
and
\be
\big[R^{(1)}\big]^{\m_1\n_1}(p) = \eta^{\m\n} \big[R^{(1)}_{\m\n}\big]^{\m_1\n_1}(p) = p^2 \eta^{\m_1\n_1} - p^{\m_1}p^{\n_1}
= p^2\,  \pi^{\m_1\n_1}(p)
\label{Ricscalp}
\ee
defined in an analogous fashion to (\ref{RieFour}). 

We also require the squared contractions 
\bea
&&\big[R_{\m\a\n\b}^{(1)}R^{(1)\m\a \n\b}\big]^{\m_1\n_1\m_2\n_2} (p_1, p_2) \equiv
\big[R_{\m\a\n\b}^{(1)}\big]^{\m_1\n_1} (p_1) \big[R^{(1)\m\a \n\b}\big]^{\m_2\n_2}(p_2) 
\nn
&&\hspace{1.5cm}= (p_1 \cdot p_2)^2\, \eta^{\m_1(\m_2}\eta^{\n_2)\n_1}
 - 2\, (p_1\cdot p_2)\, p_1\,^{\hspace{-4pt}(\m_2}\eta^{\n_2)(\n_1}p_2\,^{{\hspace{-2pt}}\m_1)}
+ p_1^{\m_2}\,p_1^{\n_2}\,p_2^{\m_1}\,p_2^{\n_1}
\label{Riemsq}
\eea
and
\vspace{-3mm}
\bea
&&\big[R_{\m\n}^{(1)}R^{(1)\m\n}\big]^{\m_1\n_1\m_2\n_2} (p_1, p_2) \equiv
\big[R_{\m\n}^{(1)}\big]^{\m_1\n_1} (p_1) \big[R^{(1)\m\n}\big]^{\m_2\n_2}(p_2) 
\nn
&&\hspace{-1.5cm}=\sdfrac{1}{4}\, p_1^2 \, \Big(p_2^{\m_1}\,  p_2^{\n_1}\, \eta^{\m_2\n_2}  
-  2\, p_2\,^{\hspace{-4pt}(\m_1}\eta^{\n_1)(\n_2}  p_2\,^{\hspace{-2.5pt}\m_2)}\Big)
+ \sdfrac{1}{4}\, p_2^2\, \Big(p_1^{\m_2}\,  p_1^{\n_2}\, \eta^{\m_1\n_1} 
-  2\,p_1\,^{\hspace{-4pt}(\m_1}\eta^{\n_1)(\n_2}  p_1\,^{\hspace{-2.5pt}\m_2)}\Big)\nn
&&+ \sdfrac{1}{4}\, p_1^2\ p_2^2\, \eta^{\m_1(\m_2}\eta^{\n_2)\n_1}
+ \sdfrac{1}{4}\, (p_1\cdot p_2)^2\, \eta^{\m_1\n_1}\eta^{\m_2\n_2} 
+ \sdfrac{1}{2} \, p_1^{(\m_1}\,p_2^{\n_1)}\,p_1^{(\m_2}\,p_2^{\n_2)}\nn
&&+  \sdfrac{1}{2} \, (p_1\cdot p_2)\,
\Big( p_1\,^{\hspace{-4pt}(\m_1}\, \eta^{\n_1)(\n_2}  p_2\,^{\hspace{-2.5pt}\m_2)}
-\eta^{\m_1\n_1}  \, p_1^{(\m_2}\,  p_2^{\n_2)} -\eta^{\m_2\n_2}  \, p_1^{(\m_1}\,  p_2^{\n_1)}\Big)\,.
\label{Riccsq}
\eea
With these expressions in hand, together with the simpler
\be
\big[(R^{(1)})^2\big]^{\m_1\n_1\m_2\n_2} (p_1, p_2) \equiv
\big[R^{(1)}\big]^{\m_1\n_1} (p_1) \big[R^{(1)}\big]^{\m_2\n_2}(p_2)
= p_1^2\, p_2^2 \, \pi^{\m_1\n_1}(p_1)\, \pi^{\m_2\n_2}(p_2)
\label{Riccscalsq}
\ee
we may express the third order anomaly action (\ref{S3anom3}) and its contribution $^{(\cA)}\cS_3$ to the 
three-point corretator in momentum space in the form
\bea
&&^{(\cA)}\cS_3^{\m_1\nu_1\m_2\nu_2\m_3\nu_3}(p_1,p_2,p_3)=\sdfrac{8}{3} \Big\{\pi^{\m_1\nu_1}(p_1)
\,\left[b'E^{(2)}+b(C^2)^{(2)}\right]^{\m_2\nu_2\m_3\nu_3}(p_2,p_3)+(\text{cyclic})\Big\}\nn
&&\qquad -\sdfrac{16b'}{9}\Big\{ \pi^{\m_1\nu_1}(p_1)\,Q^{\m_2\nu_2}(p_1,p_2,p_3)\, \pi^{\m_3\nu_3}(p_3)+(\text{cyclic})\Big\}\nn
&&\qquad \qquad +\sdfrac{16b'}{27\,}\, \pi^{\m_1\nu_1}(p_1)\,\pi^{\m_2\nu_2}(p_2)\,\pi^{\m_3\nu_3}(p_3)\,\Big\{p_3^2\, p_1\cdot p_2+(\text{cyclic})\Big\}
+ ({\rm local}) \label{AS3}
\eea
after taking account of the $2^3 = 8$ normalization factor in (\ref{Tayexp}) for $n=3$, and where the $3$ cyclic permutations of the 
indices $(1,2,3)$ are summed over. In (\ref{AS3})
\bea
&&Q^{\m_2\nu_2}(p_1,p_2,p_3) \equiv p_{1\m}\, [R^{\m\nu}]^{\m_2\nu_2}(p_2)\,p_{3\n} \nn
&&= \sdfrac{1}{2} \,\Big\{(p_1\cdot p_2)(p_2\cdot p_3)\, \eta^{\m_2\n_2} 
+ p_2^2 \ p_1^{(\m_2}\, p_3^{\n_2)} - (p_2\cdot p_3) \, p_1^{(\m_2}\, p_2^{\n_2)} - (p_1\cdot p_2) \, p_2^{(\m_2}\, p_3^{\n_2)}\Big\}
\eea
by (\ref{Riccip}), and
\bes
\bea
&&\hspace{-1.5cm}\big[E^{(2)}\big]^{\m_i\nu_i\m_j\nu_j} =\big[R_{\m\a\n\b}^{(1)}R^{(1)\m\a \n\b}\big]^{\m_i\nu_i\m_j\nu_j}
-4\,\big[R_{\m\n}^{(1)}R^{(1)\m\n}\big]^{\m_i\nu_i\m_j\nu_j}
+\big[ \big(R^{(1)}\big)^2\big]^{\m_i\nu_i\m_j\nu_j}\\
&&\hspace{-1.5cm} \big[(C^2)^{(2)}\big]^{\m_i\nu_i\m_j\nu_j}= \big[R_{\m\a\n\b}^{(1)}R^{(1)\m\a \n\b}\big]^{\m_i\nu_i\m_j\nu_j}
-2\,\big[R_{\m\n}^{(1)}R^{(1)\m\n}\big]^{\m_i\nu_i\m_j\nu_j}
+ \sdfrac{1}{3}\,\big[(R^{(1)})^2\big]^{\m_i\nu_i\m_j\nu_j}
\eea
\ees
are given by (\ref{Riemsq})-(\ref{Riccscalsq}), with the corresponding momentum dependences $(p_i, p_j)$ suppressed. The
addition term labelled `(local)' at the end of (\ref{AS3}) refers to the third variation of (\ref{bprimelocal}), the purely local last term in (\ref{S3anom3}). 

The explicit form of the contribution to the three-point vertex function $\lag TTT\rag$ from $\cS_{\rm anom}$ in (\ref{genS}) is thus (\ref{AS3}). 
We shall now verify that the contribution (\ref{AS3}) of the anomaly effective action (\ref{Sanom}) to the three-point correlator 
$\cS_3$ is precisely the same as (\ref{STrace}), obtained by the general algebraic reconstruction method of Sec. \ref{Sec:Reconst}.

We note first that (\ref{AS3}) contains no coincident double pole terms of the form $(p_i^2)^{-2}$, and it also contains no cubic single 
pole terms of the form $(p_1^2p_2^2p_3^2)^{-1}$, since the polynomial of the last term in brackets of (\ref{AS3}) contains at least 
one power of $p_1^2, p_2^2$ or $p_3^2$. However recalling (\ref{pidef}), (\ref{AS3}) does contain both single pole and quadratic single 
pole terms of the form $(p_1^2p_3^2)^{-1}$ {\it etc.} from the product of $\pi^{\m_1\nu_1}(p_1)\pi^{\m_3\nu_3}(p_3)$. It is a straightforward exercise 
in tensor algebra using (\ref{Riccip})-(\ref{Riccscalsq}) to verify that all these pole terms cancel upon taking one trace of (\ref{AS3}) so that
\bea
&&\hspace{-1.5cm}\eta_{\a_1\b_1}\,^{\!(\cA)}\cS_3^{\a_1\b_1\m_2\n_2\m_3\n_3}(p_1,p_2,p_3)\Big\vert_{p_3 = -(p_1 + p_2)} =\nn
&& 8 b\, \big[(C^2)^{(2)}\big]^{\m_2\n_2\m_3\n_3} (p_2, p_3) 
+ \,8b'\, \big[E^{(2)}\big]^{\m_2\n_2\m_3\n_3} (p_2, p_3) = 4 \,\tilde \cA_2^{\m_2\n_2\m_3\n_3} (p_2, p_3)
\label{traceS3}
\eea
yields back the second variation of the trace anomaly, consistent with the explicitly anomalous contribution to the trace
identity (\ref{threeptr}), provided again that the contribution from the local term (\ref{bprimelocal}) and $\sq R$ in the
anomaly is neglected, and account is taken of the combinatoric factor from the two possible metric variations 
of the second order $C^2$ and $E$ terms in the trace anomaly. Since
\vspace{-5mm} 
\bes
\bea
&&p_{2 \m_2} \, Q^{\m_2\n_2} (p_1, p_2, p_3) =  0\\
&&p_{2 \m_2} \, \big[(C^2)^{(2)}\big]^{\m_2\n_2\m_3\n_3} (p_2, p_3)=0\\
&&p_{2 \m_2} \, \big[E^{(2)}\big]^{\m_2\n_2\m_3\n_3} (p_2, p_3) = 0
\eea
\ees
both the second variation of the anomaly $\tilde\cA_2^{\m_2\n_2\m_3\n_3}$ and the anomaly contribution to the three-point
function $^{(\cA)}\cS_3^{\m_1\nu_1\m_2\nu_2\m_3\nu_3}$ are transverse, and hence make no contribution to the longitudinal terms
in any of the Ward Identities, as claimed in Sec.\,\ref{Sec:Reconst} and needed for the vanishing of the last term in (\ref{mixedLaTh}).

Taking an additional trace of (\ref{traceS3}), we find that the $b$ term has zero trace and there remains only
\vspace{-3mm}
\bea
&&\eta_{\a_1\b_1}\eta_{\a_3\b_3}\,^{\!(\cA)}\cS_3^{\a_1\b_1\m_2\n_2\a_3\b_3}(p_1,p_2,p_3)\big\vert_{p_3 = -(p_1 + p_2)} 
= 8b' \, \eta_{\a_3\b_3}\big[E^{(2)}\big]^{\m_2\n_2\a_3\b_3} (p_2, p_3)  \nn
&& \qquad = 16b'\, Q^{\m_2\n_2}(p_1,p_2,p_3)\big\vert_{p_3 = -(p_1 + p_2)} \, + \ 8b'\,p_2^2\, \left( p_1^2  + p_1\cdot p_2\right)\pi^{\m_2\n_2}(p_2) 
\label{dbltrace3}
\eea
in the double trace. Thus in the first line of (\ref{AS3}) we may substitute (\ref{traceS3}) and in the second line use (\ref{dbltrace3}) 
to eliminate the $Q^{\m_2\n_2}$ terms and its three cyclic permutations. Lastly computing the triple trace
\be
\eta_{\a_1\b_1}\eta_{\a_2\b_2}\eta_{\a_3\b_3}\,^{\!(\cA)}\cS_3^{\a_1\b_1\a_2\b_2\a_3\b_3}(p_1,p_2,p_3)\big\vert_{p_3 = -(p_1 + p_2)} =
16b' \left[ p_1^2\,p_2^2 - (p_1\cdot p_2)^2\right]
\label{triptrace3}
\ee
we find that we can write (\ref{AS3}) in the form
\bea
&&^{(\cA)}S_3^{\m_1\n_1\m_2\n_2\m_3\n_3}=\sdfrac{1}{3}\, \pi^{\m_1\n_1}(p_1)\,\eta_{\a_1\b_1}\,^{(\cA)}S_3^{\a_1\b_1\m_2\n_2\m_3\n_3}
+\sdfrac{1}{3}\, \pi^{\m_2\n_2}(p_2)\,\eta_{\a_2\b_2}\,^{(\cA)}S_3^{\m_1\n_1\a_2\b_2\m_3\n_3}\nn
&& +\sdfrac{1}{3}\, \pi^{\m_3\n_3}(p_3)\,\eta_{\a_3\b_3}\,^{(\cA)}S_3^{\m_1\n_1\m_2\n_2\a_3\b_3} 
-\sdfrac{1}{9}\,\pi^{\m_1\n_1}(p_1)\,\pi^{\m_3\n_3}(p_3)\,\eta_{\a_1\b_1}\eta_{\a_3\b_3}\,^{(\cA)}S_3^{\a_1\b_1\m_2\n_2\a_3\b_3} \nn
&&\hspace{-8mm}-\sdfrac{1}{9}\,\pi^{\m_2\n_2}(p_2)\pi^{\m_3\n_3}(p_3)\,\eta_{\a_2\b_2}\eta_{\a_3\b_3}\,^{(\cA)}S_3^{\m_1\n_1\a_2\b_2\a_3\b_3}
-\sdfrac{1}{9}\,\pi^{\m_1\n_1}(p_1)\pi^{\m_2\n_2}(p_2)\eta_{\a_1\b_1}\, \eta_{\a_2\b_2}\,^{(\cA)}S_3^{\a_1\b_1\a_2\b_2\m_3\n_3}\nn
&&+\sdfrac{1}{27}\,\pi^{\m_1\n_1}(p_1)\pi_2^{\m_2\n_2} (p_2)\pi^{\m_3\n_3}(p_3)\,\eta_{\a_1\b_1}\eta_{\a_2\b_2}\eta_{\a_3\b_3}\,^{(\cA)}S_3^{\a_1\b_1\a_2\b_2\a_3\b_3}\,.
\label{fin1}
\eea
Using \eqref{traceS3} once more to express all traces in terms of $\cA_2$ and comparing the result with (6.12), we see that they coincide, and thus we have proven that
\be
^{(\Th)\!}S_3^{\m_1\n_1\m_2\n_2\m_3\n_3} (p_1,p_2, p_3)\Big\vert_{\cA_2} =\, ^{(\cA)\!}S_3^{\m_1\n_1\m_2\n_2\m_3\n_3} (p_1,p_2,p_3)
\label{S3equiv}
\ee
since each is determined completely by the same single trace anomaly $\cA_2$ terms. 

The contribution from the $\sq R$ term in the trace anomaly
and variation of the local term (\ref{bprimelocal}) with arbitrary coefficient is easily included in $\cA_2$ if desired, which being local does not in any way
affect the non-local $1/p_j^2$ poles in (\ref{fin1}) or (\ref{S3equiv}). Thus the curved space anomaly effective action (\ref{Sanom}) expanded to third order 
in variations around flat space (\ref{S3anom3}) yields precisely the {\it same} anomalous trace parts and non-local propagator pole structure 
of the three-point $\lag TTT\rag$ correlation function of stress tensors for a CFT in $d=4$, obtained by the anomalous trace Ward Identities 
and general algebraic reconstruction algorithm of \cite{Bzowski:2013sza} in flat space. 

\section{Summary and Discussion}
\label{Sec:Discuss}

The principle results of this paper are Eqs.~(\ref{CWIs2}), (\ref{CWIs3a}), (\ref{CWIs3b}) for the covariant renormalized CWIs in $d = 4$; Eq.(\ref{fin1}) with (\ref{traceS3}), (\ref{dbltrace3}) and (\ref{triptrace3}) for the specifically anomalous contributions to $\lag TTT\rag$; and its complete equivalence \eqref{S3equiv} with \eqref{STraceA2} which was obtained by application of the quite different methods of Refs. \cite{Bzowski:2013sza}. In particular, exactly the same anomaly massless scalar pole is obtained from solution of the CWIs in $d = 4$ in either approach. We have emphasized that variation of the covariant action functional in a general curved space background makes possible the most compact and geometrically transparent derivation of the CWIs \eqref{CWIs2}-\eqref{CWIs3b}, including all contact terms for any CFT in $d = 4$. The general form of the exact 1PI quantum effective 
action (\ref{genS}) for a CFT in $d=4$ in the geometric approach also allows a clean
separation between the contribution to the action specifically due to the conformal anomaly, $\cS_{\rm anom}$ in any of its forms (\ref{Snonl}), (\ref{Snonlsq}) or (\ref{Sanom}), distinct from the Weyl invariant $\cS_{\rm inv}$ or local term $\cS_{\rm local}$. As a non-trivial cocycle of the local Weyl group of conformal transformations of the spacetime metric, one may always add Weyl invariant terms such as (\ref{SaddW}) to $\cS_{\rm anom}$, but such additions cannot change its conformal dependence in $\G_{_{WZ}}$ of (\ref{WZfour}), nor its pole structure in momentum space, 
required by solving (\ref{Esig}) for $\s$, incorporating the Wess-Zumino consistency condition on the covariant effective action (\ref{SanomWZ}). 
It is clear that adding covariant local terms such as (\ref{bprimelocal}) also cannot change the pole structure of the anomaly action. The presence of the scalar $D(x,x')$ propagator in \eqref{Snonl} and \eqref{Snonlsq}, and the single $(\sq)^{-1}_{xx'}$ propagator pole in \eqref{S3anom3} are irreducible necessary consequences of general covariance, the CWIs, and the conformal anomaly.

Conversely, the derivation of (\ref{Sanom}) and the decomposition (\ref{genS}) also show that the anomaly action $\cS_{\rm anom}$ certainly does not determine the Weyl invariant terms. This is a significant difference from the $d=2$ case, where {\it all} metrics are
locally conformally flat, and there are no undetermined Weyl invariant terms. In the special case of $d=2$ CFT, the non-local anomaly 
effective action is of the form $\int dx \int dy\, R(x)\,(\sq^{-1})_{xy} R(y)$ \cite{Polyakov:1981}, with $R$ the $d=2$ scalar curvature and $\sq^{-1}$ 
the Green's function inverse of the covariant wave operator, showing that the effect of the conformal anomaly involves an intermediate massless scalar 
exchange, or isolated pole in momentum space. This pole may be interpreted as that of a propagating scalar field $\vf$ which we have termed the \textit{conformalon} field,  introduced to cast the 
anomaly action in an equivalent local form. Its propagator gives rise to a massless poles in all the higher point vertex functions 
$\lag T^{\m_1\n_1}(x_1) T^{\m_2\n_2}(x_2)T^{\m_3\n_3}(x_3)...\rag_g$  of multiple energy momentum tensors, obtained by varying the effective action in $d=2$ with respect to the metric multiple times \cite{Blaschke:2014ioa}. This massless scalar exchange may be seen as an effective correlated two-particle 
$0^+$ state of the underlying quantum theory, similar to a Cooper pair of electrons in superconductivity, but appearing here in the Lorentz invariant vacuum 
state. The presence of a massless scalar pole in the three-point function (\ref{AS3}) shows that this occurs in $d=4$ as well.

What the anomaly effective action does determine are all the {\it anomalous} contributions to the higher point stress tensor
correlation functions, since the exact 1PI quantum action (\ref{genS}) is precisely the generating function for these connected
correlation functions in an arbitrary fixed background, and $\cS_{\rm anom}$ is the only term responsible for anomalous 
contributions in the trace Ward Identities. In this paper we have calculated the contribution of $\cS_{\rm anom}$
to the three-point function, given explicitly by (\ref{AS3}), by three variations of the general curved space anomaly action.
The equality (\ref{S3equiv}) shows that this is {\it precisely the same result} as that obtained for the anomalous trace parts
by the method of \cite{Bzowski:2013sza} of solving first the exact CWIs for the projected transverse, traceless parts of the
correlator directly in $d$-dimensional flat space, then reconstructing the full $\lag TTT\rag$ by restoring the longitudinal and 
trace parts by use of the conservation and anomalous trace Ward Identities in the $d\rightarrow 4$ limit. This is an explicit verification 
of $\cS_{\rm anom}$ for the anomalous trace parts of the full $\lag TTT\rag$ in any $d=4$ CFT, including the presence of multiple pole 
terms in all the external invariants, predicted by the anomaly action. In the approach of \cite{Bzowski:2013sza} these 
multiple pole contributions are just a consequence of the transverse projection operators (\ref{pidef}) in the reconstruction 
formula (\ref{STrace}) for the trace parts of the three-point function. These trace contributions, in complete 
agreement with the variation of the anomaly action, have the correct analytic structure to satisfy the {\it anomalous} CWIs of a CFT
in the physical $d=4$ dimensions.

It is clear from this derivation and the equivalence (\ref{S3equiv}) that the massless pole contributions are unambiguously fixed and 
determined by the anomaly effective action, as a necessary consequence of the anomalous trace and conformal Ward Identities. 
That anomalies are generally associated with massless poles in $d=4$ has taken some time to recognize, although the prototype example 
was already provided by the $d=2$ Schwinger model and Polyakov action decades ago \cite{Blaschke:2014ioa}. Massless poles in $\lag TTT\rag$ imply that there are long range effects of stress tensor correlations on lightlike separated 
spacetime points \cite{Giannotti:2008cv,Armillis:2009pq}, similar to those already noted in the case of the chiral anomaly 
\cite{Armillis:2009sm,Coriano:2008pg,Giannotti:2008cv}. In the supersymmetric case this behavior is present in all the components 
of a superconformal anomaly multiplet \cite{Coriano:2014gja}. This light cone behavior is most clearly seen in a Lorentzian
momentum space representation.

Because of the coupling 
to $T^{\m\n}$, massless poles on the light cone lead to novel scalar gravitational effects on macroscopic scales, 
not described by Einstein's classical theory \cite{AndMolMott:2003, Mottola:2006ew,Mottola:2016mpl,Meissner:2016onk}. 
In particular, the appearance of a propagating effective massless scalar degree of freedom in $d=4$, explicitly represented by the local 
conformalon field $\vf$ in (\ref{Sanom}) has implications for gravity at low energies and at macroscopic scales \cite{Mottola:2006ew}, 
including the existence and propagation of scalar gravitational waves not present in classical General Relativity \cite{Mottola:2016mpl}. Further implications of this effective light scalar in the Effective Field Theory of four dimensional gravity and propagator poles in higher point stress tensor correlation functions are under investigation.

\vfil\break
\centerline{\bf Acknowledgements}
\vspace{5mm}

C. C. and E. M. thank A. Bzowski, P. McFadden, and K. Skenderis for discussions and clarifications on their work, and especially K. Skenderis 
for invitations to the Univ. of Southampton where the present work was begun. E.M. thanks H. Osborn for his comments on this work and its agreement with Ref.\cite{Erdmenger:1996yc}, and also thank G. Goon for helping to clarify
the terms properly identified in Eq. (\ref{S3equiv}). The work of C.C. was supported in part by a {\em The Leverhulme 
Trust Visiting Professorship} at the STAG Research Centre and Mathematical Sciences, University of Southampton. C. C. also acknowledges 
helpful discussions with M. Serino and L. Delle Rose. The work of C.C. is partially supported by INFN, iniziativa specifica HEP-QFT.


\end{document}